\documentclass[journal]{IEEEtran}
\IEEEoverridecommandlockouts
\usepackage{cite}
\usepackage{amsmath,amssymb,amsfonts}
\usepackage{graphicx}
\usepackage{textcomp}
\usepackage{xcolor}
\usepackage{multirow}
\usepackage{makecell}
\usepackage{algorithm,algcompatible}

\def\BibTeX{{\rm B\kern-.05em{\sc i\kern-.025em b}\kern-.08em
    T\kern-.1667em\lower.7ex\hbox{E}\kern-.125emX}}

\long\def\comment#1{}

\newfont{\bbb}{msbm10 scaled 700}

\newfont{\bb}{msbm10 scaled 1100}
\newcommand{\CC}{\mbox{\bb C}}

\newcommand{\EE}{\mbox{\bb E}}

\newcommand{\av}{{\bf a}}
\newcommand{\bv}{{\bf b}}

\newcommand{\hv}{{\bf h}}

\newcommand{\nv}{{\bf n}}

\newcommand{\tv}{{\bf t}}

\newcommand{\vv}{{\bf v}}
\newcommand{\xv}{{\bf x}}
\newcommand{\yv}{{\bf y}}
\newcommand{\zv}{{\bf z}}

\newcommand{\Am}{{\bf A}}

\newcommand{\Cm}{{\bf C}}
\newcommand{\Dm}{{\bf D}}

\newcommand{\Fm}{{\bf F}}
\newcommand{\Gm}{{\bf G}}
\newcommand{\Hm}{{\bf H}}
\newcommand{\Id}{{\bf I}}

\newcommand{\Mm}{{\bf M}}
\newcommand{\Nm}{{\bf N}}

\newcommand{\Pm}{{\bf P}}
\newcommand{\Qm}{{\bf Q}}
\newcommand{\Rm}{{\bf R}}
\newcommand{\Sm}{{\bf S}}
\newcommand{\Tm}{{\bf T}}
\newcommand{\Um}{{\bf U}}

\newcommand{\Vm}{{\bf V}}
\newcommand{\Xm}{{\bf X}}
\newcommand{\Ym}{{\bf Y}}
\newcommand{\Zm}{{\bf Z}}

\newcommand{\Ic}{{\cal I}}

\newcommand{\Lc}{{\cal L}}

\newcommand{\nuv}{\hbox{\boldmath$\nu$}}

\newcommand{\thetav}{\hbox{\boldmath$\theta$}}

\newcommand{\Sigmam}{\hbox{\boldmath$\Sigma$}}
\newcommand{\Phim}{\hbox{\boldmath$\Phi$}}

\newcommand{\Psim}{\hbox{\boldmath$\Psi$}}
\newcommand{\Thetam}{\hbox{\boldmath$\Theta$}}

\newcommand{\SNR}{{\sf SNR}}

\newcommand{\eqdef}{\stackrel{\Delta}{=}}

\newcommand{\herm}{{\sf H}}

\newcommand{\RED}{\color[rgb]{1.00,0.10,0.10}}
\newcommand{\BLUE}{\color[rgb]{0,0,0.90}}

\DeclareMathOperator*{\argmin}{arg\,min}

\newtheorem{remark}{Remark}

\newtheorem{definition}{Definition}

\begin{document}

\title{Near-Field Channel Estimation for XL-RIS Assisted Multi-User XL-MIMO Systems:\\ Hybrid Beamforming Architectures}

\author{Jeongjae~Lee,~\IEEEmembership{Student Member,~IEEE}, Hyeongjin~Chung,~\IEEEmembership{Student Member,~IEEE}, Yunseong~Cho,~\IEEEmembership{Member,~IEEE}, Sunwoo~Kim,~\IEEEmembership{Senior Member,~IEEE}
        and~Songnam~Hong,~\IEEEmembership{Member,~IEEE}% <-this % stops a space
\thanks{J. Lee, H. Chung, S. Kim, and S. Hong are with the Department of Electronic Engineering, Hanyang University, Seoul, Korea (e-mail: \{lyjcje7466, hyeonjingo, remero, snhong\}@hanyang.ac.kr).}
\thanks{Y. Cho is with Samsung Research America, Plano, Texas, USA (e-mail: yscho@utexas.edu).}
% <-this % stops a space Yunseong~Cho,~\IEEEmembership{Member,~IEEE}
%\thanks{This paper was presented in part at the 2020 IEEE International Symposium on Information Theory.}
%Y. Cho is with Samsung Research America, Plano, Texas, USA. email: yscho@utexas.edu

\thanks{This work was supported in part by the Technology Innovation Program (1415178807, Development of Industrial Intelligent Technology for Manufacturing, Process, and Logistics) funded By the Ministry of Trade, Industry \& Energy(MOTIE, Korea) and in part by the Institute of Information \& communications Technology Planning \& Evaluation (IITP) under the artificial intelligence semiconductor support program to nurture the best talents (IITP-(2024)-RS-2023-00253914) grant funded by the Korea government(MSIT).}}

\maketitle

\begin{abstract}
Channel estimation is one of the key challenges for the deployment of extremely large-scale reconfigurable intelligent surface (XL-RIS) assisted multiple-input multiple-output (MIMO) systems. In this paper, we study the channel estimation problem for XL-RIS assisted multi-user XL-MIMO systems with hybrid beamforming structures. For this system, we propose an {\em unified} channel estimation method that yields a notable estimation accuracy in the near-field BS-RIS and near-field RIS-User channels (in short, near-near field channels), far-near field channels, and far-far field channels. Our key idea is that the effective (or cascaded) channels to be estimated can be each factorized as the product of low-rank matrices (i.e., the product of the common (or user-independent) matrix and the user-specific coefficient matrix). The common matrix whose columns are the basis of the column space of the BS-RIS channel matrix is efficiently estimated via a {\em collaborative} low-rank approximation (CLRA). Leveraging the hybrid beamforming structures, we develop an efficient iterative algorithm that jointly optimizes the user-specific coefficient matrices. Via experiments and complexity analysis, we verify the effectiveness of the proposed channel estimation method (named CLRA-JO) in the aforementioned three classes of wireless channels.
\end{abstract}


%% 수정 중
%%%%%%%%%%%%%%%%%%%%%%%%%%%%%%%%%%%%%%%
\section{Introduction}\label{sec:Intro}

Reconfigurable intelligent surface (RIS) is a promising technology for robust millimeter-wave (mmWave) and terahertz (THz) multiple-input multiple-output (MIMO) systems \cite{di2020smart, pei2021ris, wu2021intelligent}. A RIS consists of a uniform array with a large number of reflective elements, each of which can control the phase and reflection angle of the incident signal so that the received power of the intended signal is enhanced \cite{long2020active}. The potential advantages of the RIS open up new research opportunities such as reflect beamforming design \cite{zhou2020robust, zhou2021stochastic, wu2019intelligent, liao2023optimized} and RIS-aided localization and sensing \cite{win2022location, wang2022location, wymeersch2020radio}. Nonetheless, the accuracy of a channel estimation plays a crucial role in implementing numerous RIS-assisted applications
\cite{wymeersch2020radio}.

%Existing WORKS (mmWave MIMO channel)

There have been numerous works on the design of an efficient channel estimation method in RIS-aided mmWave MIMO systems. Herein, the channel responses between the base station (BS) and the RIS (in short, the BS-RIS channel) and between the RIS and the users (in short, the RIS-User channels) are assumed as {\em far-field} channels. Accordingly, both BS and RIS will experience {\em planar-wavefront}. The primary goal of the existing works is to estimate the channel accurately while having an affordable training overhead. Toward this, the key idea is to exploit the {\em sparsity} of mmWave channels, which comes from the fact that there exists a smaller number of signal paths in the BS-RIS channel. The de-facto channel estimation method to harness the sparsity is based on compressed sensing (CS). Specifically, the angle-of-departures (AoDs) and the angle-of-arrivals (AoAs) of sparse signal paths are estimated via CS, thus requiring less training overhead. In \cite{tsai2018efficient}, the popular orthogonal matching pursuit (OMP) was used as a low-complexity CS method. An enhanced CS-based method was developed in \cite{chen2023channel}, by exploiting the subspace estimation and the so-called scaling property of the RIS-User channels. Unfortunately, the CS-based methods suffer from the inevitable {\em grid-mismatch} problem \cite{chi2011sensitivity} since a dictionary matrix is constructed by quantizing steering vectors at a specific resolution. Having an affordable computational complexity (i.e., the quantization levels), the CS-based methods can result in a severe error-floor problem due to the quantization errors.

%additionally harnessing the sparsity of the RIS-User channels. 
Besides the CS-based channel estimation methods, approximate message passing (AMP)-based methods were developed in \cite{wei2021channel, guo2022efficient}, which is based on the parallel factor decomposition to unfold the so-called effective (or cascaded) channels to be estimated. In \cite{wei2021channel}, alternating least squares (ALS) and vector AMP (VAMP) were presented as efficient iterative algorithms. Also, in \cite{guo2022efficient}, unitary AMP (UAMP)-based method was presented, which is more computationally efficient than the ALS- and VAMP-based methods. Noticeably, the AMP-based methods are only applicable in larger systems such that the number of users (denoted by $K$) is not less than the number of reflective elements (denoted by $N$) in the RIS. In the example of $N=128$, the AMP-based methods can be used at least when $K \geq 128$. This can limit the applicability of the AMP-based channel estimation methods in various RIS-aided applications.

%%%% X-MIMO %%%%%

In future communication systems (e.g., 6G), the number of reflective elements in the RIS tends to be extremely large (XL) to further enhance the beamforming gain \cite{Cui2022, liu2023low}. This will lead to a larger Rayleigh distance (RD), which is the boundary between the near- and far-field channels. For example, when the RIS is equipped with 256 half-wavelength spacing elements at operating 100GHz, the RD becomes almost $100{\rm m}$ \cite{Rapp2019}. In the 6G systems, thus, the near-field effect becomes non-negligible and accordingly, the RIS will experience {\em spherical-wavefront} rather than planar-wavefront (see Fig. 2(b)). Recall that the aforementioned channel estimation methods based on the planar-wavefront heavily rely on the channel sparsity in the angular domain. Due to the near-field effect in the spherical-wavefront, however, the channel sparsity in the angular domain is no longer applicable. It is necessary to develop an efficient channel estimation method for near-field XL-RIS assisted MIMO systems. Assuming that the RIS-User channels are near-field, the channel estimation problem was formulated as the CS problem by means of the polar-domain sparsity \cite{Cui2022, Lu2023}. Based on this, a single measurement vector (SMV)-based OMP algorithm was presented as an efficient channel estimation method. As expected, this method suffers from the grid-mismatch problem.

In the above works, it is assumed that BS is equipped with fully-digital structures. However, when the BS is equipped with a massive or extremely large number of antennas, the use of a large number of radio frequency (RF) chains (i.e., one for each antenna component) significantly increases the cost and the energy consumption \cite{lin2016terahertz, molisch2017hybrid}. A promising solution to these problems lies in the concept of hybrid beamforming structures, which takes the combination of analog beamforming in the RF domain, together with digital beamforming in the baseband connected to the RF chains \cite{molisch2017hybrid}. Therefore, XL-RIS assisted XL-MIMO systems with hybrid beamforming structures can be considered as a promising energy-efficient technology \cite{wang2020joint, yildirim2022ris, 9743307}. Recently in \cite{schroeder2022channel,chung2023location,JJLee2023, chung2024efficient}, the channel estimation method for RIS-aided mmWave MIMO systems with hybrid beamforming structures was investigated, in which all channels are assumed as {\em far-field} channels. In \cite{schroeder2022channel, chung2023location}, the grid-mismatch problem was alleviated via atomic norm minimization (ANM). However, the ANM-based methods are impractical due to their expensive computational complexities. In \cite{JJLee2023,chung2024efficient}, the channel estimation problem was formulated as a low-rank matrix completion (LRMC) and then solved via fast alternating least squares (FALS) \cite{hastie2015matrix}. Also, it was shown that the LRMC-based method can outperform the CS-based methods by avoiding the grid-mismatch problem \cite{chung2024efficient, JJLee2023}. Nevertheless, this method suffers from the {\em noisy-sample} problem, which is inevitable due to the use of matrix completion with {\em noisy} observations. Specifically, the sampled noisy elements cannot be updated via the proposed matrix completion method in \cite{chung2024efficient, JJLee2023} and thus, the estimation accuracy is saturated as the pilot overhead (i.e., the number of sampled elements) grows. In \cite{Yang2023,Yu2023}, the XL-RIS assisted MU-MIMO systems with hybrid beamforming structures, which is the most related to our system model, was investigated, where the BS-RIS and the RIS-User channels are assumed as {\em far- and near-field} channels, respectively. In \cite{Yang2023}, the two-phase channel estimation protocol was proposed to improve the efficiency of training overhead. Based on this framework, a 3-dimensional multiple measurement vector (MMV) look-ahead OMP (3D-MLAOMP) was developed as an efficient channel estimation method. This method still suffers from the grid-mismatch problem and requires some impractical assumptions such as the number of spatial paths and the angle of line-of-sight (LoS) path being given as a priori. In \cite{Yu2023}, a fast-sparse-Bayesian-learning (FSBL)-based channel estimation method was proposed by focusing on the limited visual region (VR) of the channel between the RIS and each user in XL-RIS systems. This method is impractical due to the expensive computational complexity since a super-resolution (SR) algorithm is used to estimate the common AoD at the BS and its complexity is too expensive. Thus, it is an open problem to develop an efficient channel estimation method for XL-RIS assisted XL-MIMO systems.
%3D-MLAOMP의 경우는 XL-RIS는 near-field, XL-MIMO는 far-field를 고려했습니다.

%some assumptions의 경우는 
%1) first phase의 결과를 second phase에 반영할 때 steering vector의 asymptotic orthogonality를 적용한다는 점
%2) RIS/UE side dictionary를 디자인할 때 (near-field와 far-field가 혼합되어 있는 부분), LoS path를 가지고 디자인을 하는데, 이때 LoS path에 해당하는 AoD, AoA를 알고 있어야 한다는 점

%We consider that this XL-RIS-aided MIMO systems can be extended to the general RIS-aided XL-MIMO systems, where both the BS and the RIS are equipped with extremely large-scale antenna array (ELAA) and the channel can be modeled with XL-MIMO and XL-MISO channel proposed in \cite{Lu2023} considering the MIMO advanced Rayleigh distance (MIMO-ARD). In this system, the channel between the BS and the RIS can be also a near-field channel, where the aforementioned works become limited to utilize for channel estimation.

Motivated by the above, in this paper, we study the channel estimation problem for XL-RIS assisted multi-user XL-MIMO systems with hybrid beamforming structures. Noticeably, it is assumed that both the BS and the RIS are equipped with extremely large-scale antenna array (ELAA). Accordingly, we consider the following categories of wireless channels: i) Far-field BS-RIS and far-field RIS-User channels (i.e., {\em far-far} field channel); ii) Far-field BS-RIS and near-field RIS-User channels (i.e., {\em far-near} field channel); iii) Near-field BS-RIS and near-field RIS-User channels (i.e., {\em near-near} field channel). Beyond the existing works in \cite{Yu2023,Yang2023}, we for the first time investigate the channel estimation method for near-near field channels. In addition, as an extension of the existing works, the proposed channel estimation method can be performed when each user is equipped with a multiple antenna.
%Motivated by the above, in this paper, we study the channel estimation problem for RIS-aided mmWave MU-MIMO systems with hybrid beamforming structures (see Fig. 1). In addition, as an extended system of the related works in \cite{chung2022efficient, JJLee2023}, it is assumed that each user is equipped with a multiple transmit antenna. 
The major contributions of this paper are summarized as follows.

\begin{itemize}
   \item We develop an {\em unified} channel estimation method that can be performed on {\em near- and far-field} uniform planar array (UPA) MIMO channels without any modification. Furthermore, we present a two-phase communication protocol suitable for the proposed channel estimation method, by harnessing the fact that BS-RIS and RIS-User channels have different channel coherence times. This can considerably reduce the training overhead.

    \item In mmWave and THz communication systems, there exists a small number of signal paths in the BS-RIS channel. Thus, the effective (or cascaded) channels to be estimated (denoted by  $\{\Hm_{[{\rm eff},k]}: k \in [K]\}$) have lower-dimensional {\em common} column space regardless of the aforementioned channel categories. Leveraging this, each effective channel can be categorized as the product of low-rank matrices, i.e.,  $\Hm_{[{\rm eff},k]} = \Sm_{\rm col}\Tm_{k}$, where $\Sm_{\rm col}$ contains the basis of the common column space and $\Tm_{k}$ denotes a user-specific coefficient matrix.

   \item Based on the above low-rank factorization, the proposed channel estimation method consists of two parts. In the first part, we estimate the dimension of the column space (i.e., the size of the matrix $\Sm_{\rm col}$)) using the minimum description length (MDL) criterion and then estimate $\Sm_{\rm col}$ by means of a collaborative low-rank approximation (CLRA). Via theoretical analysis, it is proved that the accuracy of the estimated $\Sm_{\rm col}$ is inversely proportional to the product of the pilot overhead and the number of users $K$. To achieve a target accuracy level, the pilot overhead can be reduced as the pilot overhead grows, namely, a multi-user gain can be attained.

   \item In the second part, we propose an efficient iterative algorithm to {\em jointly} optimize the user-specific coefficient matrices $\{\Tm_{k}: k \in [K]\}$. Herein, the joint optimization is formulated using the fact that $\Hm_{[{\rm eff},k]}=\Dm_{k}\Hm_{[{\rm eff},1]}$, $\forall k \in [K]$, with some diagonal matrix $\Dm_k$. We show that this joint optimization can significantly enhance the signal-to-noise ratio (SNR) gain compared with the standard {\em individual} least-square (LS) estimations. Also, the convergence of the proposed iterative algorithm is theoretically proved.

   \item The proposed channel estimation method is referred to as {\bf C}ollaborative {\bf L}ow-{\bf R}ank {\bf A}pproximation and {\bf J}oint {\bf O}ptimization (CLRA-JO). Via experiments and complexity analysis, we verify the effectiveness of the proposed CLRA-JO. In various channel environments, the proposed method can yield better estimation accuracy than the state-of-the-art CS-based methods while having lower training overhead (e.g., about $80\%$ reduction of the training overhead).
\end{itemize}

The remaining part of this paper is organized as follows. In Section II, we define the channel and signal models for XL-RIS assisted multi-user XL-MIMO systems with hybrid beamforming structures. Section III introduces the channel estimation protocol and the frame structures. In Section IV, we describe the proposed channel estimation method, named CLRA-JO. Section V provides simulation results and Section VI concludes the paper.

%%%%%%%%%%%%%%%%%%%%%%%%%%%%%%%%%%%%%%%%%%%%%%%%%%%%%%%%%%%%%%%
{\em Notations.} Let $[N_1:N_2]\eqdef\{N_1,N_1+1,...,N_2\}$ for any positive integers $N_1$ and $N_2$ with $N_1 < N_2$. When $N_1=1$, it is further simplified as $[N_2]\eqdef\{1,2,...,N_2\}$. We use $\xv$ and $\Am$ to denote a column vector and matrix, respectively. Also, $\Am^{\dagger}$ denotes the Moore-Penrose inverse and $\otimes$ denotes the Kronecker product. Given a $M \times N$ matrix $\Am$, let $\Am(i,:)$ and $\Am(:,j)$ denote the $i$-th row and $j$-th column of $\Am$, respectively, Also, given the index subsets $\Ic_{\rm row}\subseteq [M]$ and $\Ic_{\rm col} \subseteq [N]$, we let $\Am(\Ic_{\rm row},:)$ and $\Am(:,\Ic_{\rm col})$ denote the submatrices of $\Am$ by only taking the rows and columns whose indices are belong to $\Ic_{\rm row}$ and $\Ic_{\rm col}$, respectively. Given a vector $\vv$, ${\rm diag}(\vv)$ denotes a diagonal matrix whose $\ell$-th diagonal element is equal to the $\ell$-th element of $\vv$. We let $\Id$ and ${\bf 0}$ denote the identity and all-zero matrices, respectively, where the sizes of these matrices are easily obtained from the context.

%%%%%%%%%%%%%%%%%%%%%%%%%%%%%%%%%%%%%%
\begin{figure}[t]
\centering
\includegraphics[width=1.0\linewidth]{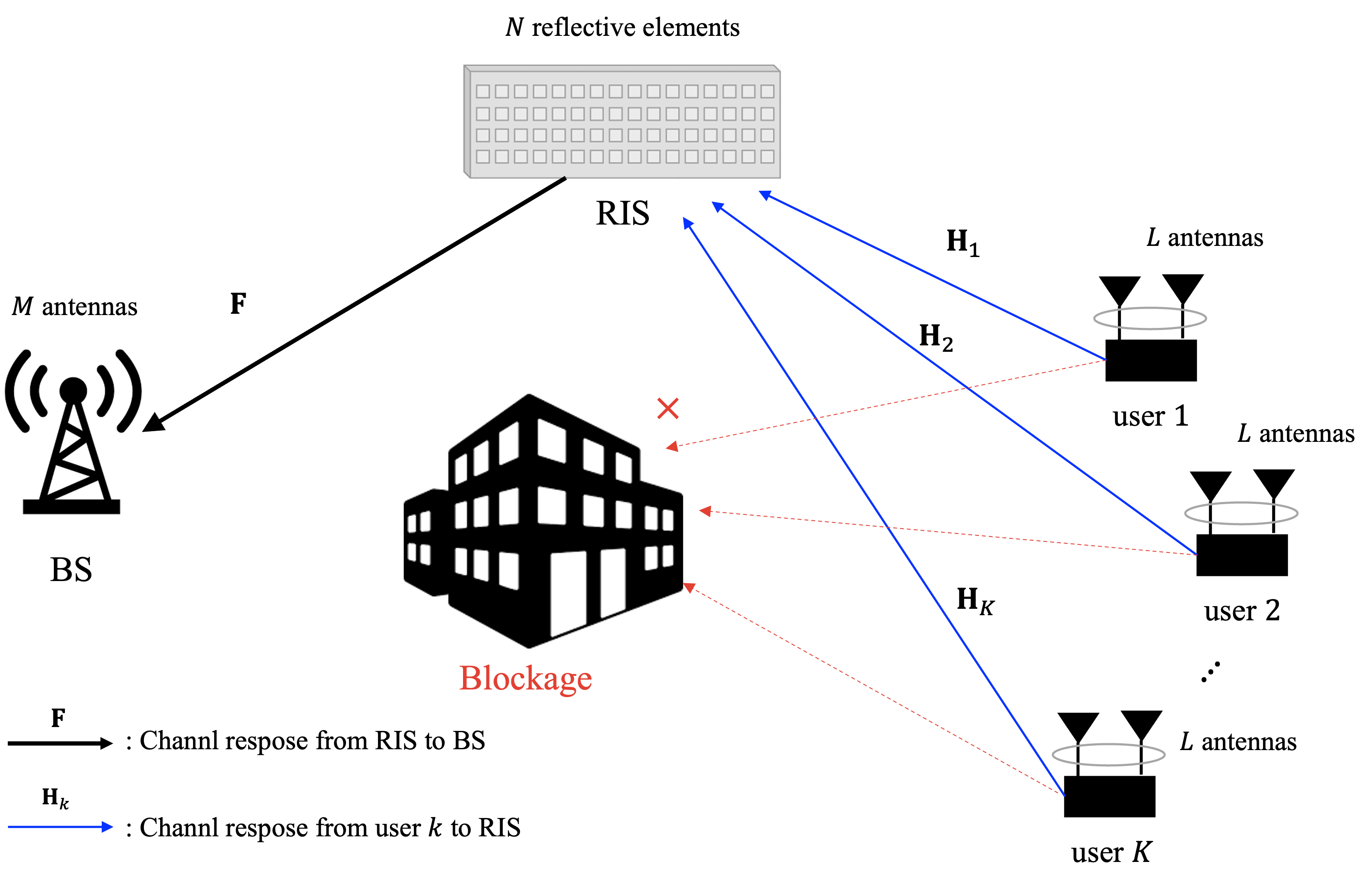}
\caption{A RIS-aided MU-MIMO system consisting of the BS with $M$ antennas, the RIS with $N$ reflective elements, and the $K$ users with $L$ antennas.}
\label{fig:system_model}
\end{figure}

%아래 REFERENCE 추가
%%%%%%%%%%%%%%%%%%%%%%%%%%%%%%%%%%%%%%%%%%%%%
% System Model
%%%%%%%%%%%%%%%%%%%%%%%%%%%%%%%%%%%%%%%%%%%%%
\section{System Model}\label{sec:System Model}

We consider an uplink time-division-duplexing (TDD) multi-user multiple-input multiple-output (MU-MIMO) system. In this system, a reconfigurable intelligent surface (RIS) with $N$ reflective elements assists the communication between the base station (BS) with $M$ antennas and the $K$ users each having $L$ antennas (see Fig.~\ref{fig:system_model}). Noticeably, it is assumed that both the BS and the RIS are equipped with extremely large-scale antenna array (ELAA). For the sake of lower complexity, cost and power consumption \cite{ahmed2018survey, molisch2017hybrid}, the BS is assumed to use the hybrid beamforming with a limited number of radio frequency (RF) chains, where $N_{\rm RF}\leq M$ denotes the number of RF chains at the BS. As considered in the related works \cite{JJLee2023,chung2024efficient ,Yu2023,Yang2023}, it is assumed that the direct channels between the BS and the $K$ users are blocked (see Fig.~\ref{fig:system_model}). The channel responses from the $k$-th user to the RIS and from the RIS to the BS are respectively denoted as  $\Hm_{k}\in\CC^{N\times L}$ and $\Fm\in\CC^{M\times N}$. The reflection vector in the RIS is denoted as $\vv=[v_1,v_2,...,v_N]^{\herm}$ with $v_{n} = e^{j\vartheta_{n}}$, where $\vartheta_{n}\in[0,2\pi)$ is the phase shift of the $n$-th reflective element. We note that the proposed channel estimation method in Section~\ref{sec:method} can be applied to both continuous and discrete phase shifting RISs. Taking this into account,  the total channel response from the $k$-th user to the BS is defined as
\begin{align}
    \Fm {\rm diag}(\vv)\Hm_{k}&=\Hm_{[{\rm eff},k]}\left(\Id\otimes\vv\right),\label{eq:channel_model}
\end{align} where the effective (or cascaded) channel is defined as
\begin{align}
    \Hm_{[{\rm eff},k]}&\eqdef\begin{bmatrix}
        \Hm_{[{\rm eff},k,1]} & \cdots & \Hm_{[{\rm eff},k,L]}
    \end{bmatrix}\in\CC^{M \times NL},\label{eq:channel_model1}
\end{align} and where 
\begin{align}
    &\Hm_{[{\rm eff},k,\ell]} =\Fm {\rm diag}\left(\Hm_{k}(:,\ell)\right)\in\CC^{M \times N}.\label{eq:channel_model2} 
\end{align} 

\begin{remark}\label{remark:effective_channel}
    We emphasize that the definition of our effective channel in \eqref{eq:channel_model1} is different from that in \cite{schroeder2022channel, chung2023location}. In the latter, an effective channel is defined as the Kronecker product of the BS-RIS and RIS-User channels. This expression is not suitable for the joint optimization of the reflection vector and the hybrid beamforming due to its large dimension. In contrast, the proposed effective channel has a more compact form, thereby being more adequate for the joint optimization, as shown in \cite{Li2022, Wu2020,Moon2022,lee2024asymptotically}. Motivated by this, we in this paper aim at developing an efficient method to estimate the effective channels $\{\Hm_{[{\rm eff},k]}:k\in[K]\}$. \hfill$\blacksquare$
\end{remark}
\vspace{0.1cm}

    %%%%%%%%%%%%%%%%%%%%%%%%%%%%%
%\renewcommand{\arraystretch}{1.0} % Default value: 1
%\begin{table}
%{\caption{Summary of notations}} 
%\centering
%\begin{tabular}{ c c  } 
%\Xhline{2\arrayrulewidth}
%\\
%$M$ & The number of receive antennas at the BS\\
%$N_{\rm RF}$& The number of RF chains at the BS\\
%%$M_{\rm RF}=\frac{M}{N_{\rm RF}}$& Assumed to be integer\\
%$N$ & The number of reflective elements at the RIS \\ 
%$K$ & The number of users\\
%$L$ & The number of transmit antennas at each user\\
%$N_f$ & The number of spatial paths between the BS and the RIS\\
%$N_{h_k}$ & The number of spatial paths between the RIS and the user $k$\\
%\\
%\Xhline{2\arrayrulewidth}
%\end{tabular}
%\end{table}
%%%%%%%%%%%%%%%%%%%%%%%%%%%%%%%%%%%%%%%%%%%%%%%%%%%%%%%%%%%%%%%%%%%%%%%%
\begin{figure*}
\centering
\includegraphics[width=1.0\linewidth]{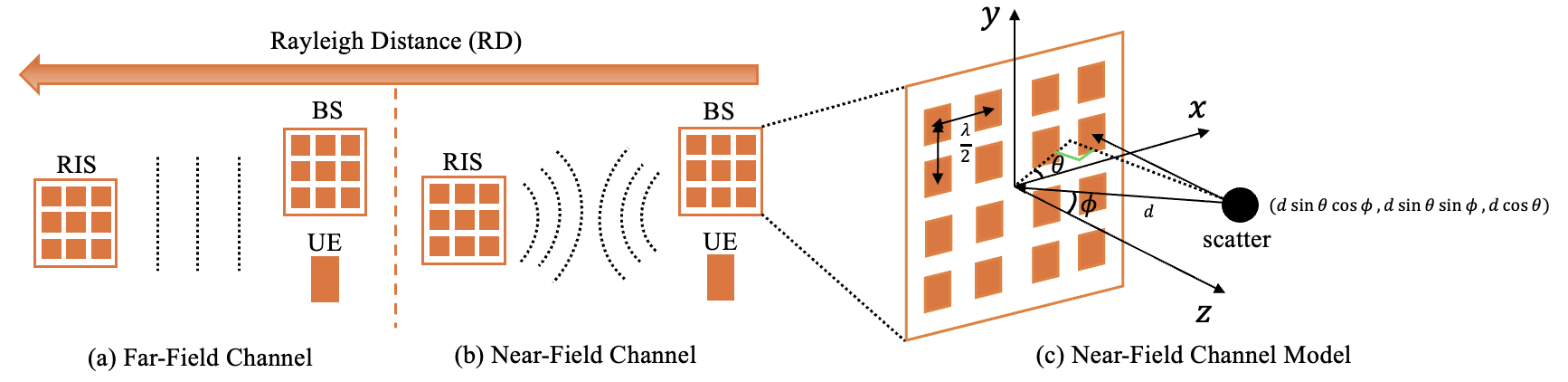}
\caption{The categories of wireless channels $\{\Hm_k:k\in[K]\}$ and $\Fm$. (a) The far-field region, where the planar-wavefront is assumed. (b) The near-field region, where the spherical-wavefront is assumed. (c) The near-field channel model for UPA communication system based on 3D-coordinate.}
\label{fig:model}
\end{figure*}
%%%%%%%%%%%%%%%%%%%%%%%%%%%%%%%%%%%%%%%%%%%%%%%%%%%%%%%%%%%

In the remaining part of this section, we will describe the specific channel models for $\{\Hm_k:k\in[K]\}$ and $\Fm$. We first provide the useful definition:
\begin{definition}\label{def:array_vector}
Given an array size $ X = E_h\times E_v$, a distance between the scatter and the  center of array $d$, an elevation angle $\theta\in[0,2\pi)$ and azimuth angle $\phi\in[0,\pi)$, we define a {\em near-field} array response vector \cite{zheng2023extremely} (see Fig. 2 (c)):
\begin{equation}
    \bv_{X}(d,\theta,\phi) \eqdef \sqrt{\frac{1}{X}}\left[e^{-j\frac{2\pi}{\lambda}\Delta d(1,1)},\dots, e^{-j\frac{2\pi}{\lambda}\Delta d(E_h,E_v)}\right]^{\intercal},
\end{equation} where for $e_h \in [E_h], e_v \in [E_v]$,
\begin{align}
    \Delta d(e_h,e_v) &\approx - \sin\theta\cos\phi x_{e_h} - \sin\theta\sin\phi y_{e_v}\nonumber\\
    &+\frac{1-\sin^{2}\theta\cos^{2}\phi}{2d}x_{e_h}^2 + \frac{1-\sin^{2}\theta\sin^{2}\phi}{2d}y_{e_v}^2\nonumber\\
    &-\frac{\sin^2\theta\cos\phi\sin\phi}{d}x_{e_h}y_{e_v}, \label{eq:UPAnear}
    %\zeta(e_h,e_v) - d
\end{align} where $x_{e_h} = \frac{\lambda}{2}(e_h-(E_h+1)/2)$ and $y_{e_v} = \frac{\lambda}{2}(e_v-(E_v+1)/2)$.
%Here, $\zeta(e_h,e_v)$ denotes the distance between the $(e_h,e_v)$-th antenna with the position $(x_{e_h} = \frac{\lambda}{2}(e_h - (E_h+1)/2), y_{e_v}=\frac{\lambda}{2}(e_v - (E_v+1)/2), 0)$ and the scatter with the position $(d\sin\theta\cos\phi, d\sin\theta\sin\phi, d\cos\theta)$. According to the second order Taylor series expansion, it can be well-approximated as 
%\begin{align}
%    \zeta(e_h, e_v) =&\; d - \sin\theta\cos\phi x_{e_h} - \sin\theta\sin\phi y_{e_v}\nonumber\\
%    &+\frac{1-\sin^{2}\theta\cos^{2}\phi}{2d}x_{e_h}^2 + \frac{1-\sin^{2}\theta\sin^{2}\phi}{2d}y_{e_v}^2\nonumber\\
%    &-\frac{\sin^2\theta\cos\phi\sin\phi}{d}x_{e_h}y_{e_v}. \label{eq:UPAnear}
%\end{align} 
When $d$ is sufficiently large, the last three terms in \eqref{eq:UPAnear} can be negligible. In this case, the array response vector $\bv_{X}(d,\theta,\phi)$ is simplified as
\begin{align}
    \av_{X}(\theta,\phi) &\eqdef \sqrt{\frac{1}{X}}\left[1,\dots, e^{j\pi(e_h\sin\theta\cos\phi+e_v\sin\theta\sin\phi)},\dots,\right.\nonumber\\
    &\;\;\;\;\;\;\;\left.e^{j\pi((E_h-1)\sin\theta\cos\phi+(E_v-1)\sin\theta\sin\phi)}\right]^{\intercal}.
\end{align} Note that this is equivalent to the conventional {\em far-field} UPA array response vector. \hfill$\blacksquare$
\end{definition}
\vspace{0.1cm}

In Section~\ref{subsec:URC} and Section~\ref{subsubsec:BS-RIS}, we describe the specific wireless channel models according to the distances between the BS and the RIS, and between the RIS and the $K$ users. Throughout the paper, we let $z_f$ and $z_{h_k}$ denote the physical distances between the BS and the RIS, and between the RIS and the user $k$, respectively.

%%%%%%%%%%%%%%%%%%%%%%%%%%%%%%%%%%%%%%%%%%%%%%%%
\subsection{RIS-User Channels}\label{subsec:URC}

We describe the RIS-User channels (i.e., $\Hm_{k}$ for $k \in [K]$). To categorize these channels, we define the Rayleigh distance (RD), given by \cite{Cui2022}:
\begin{equation}
    Z_{\rm RD} = \frac{2D_{\rm RIS}^2}{\lambda},
\end{equation} where $\lambda$ is the wavelength of EM wave and $D_{\rm RIS}$ is the RIS aperture (i.e., $D_{\rm RIS}=(\lambda/2)N$). As depicted in Fig.~\ref{fig:model}, each $\Hm_k$ can be classified into the near- and far-field channels. Specifically, for $z_{h_k} < Z_{\rm RD}$, $\Hm_{k}$ is considered as the near-field channel and  the wavefronts are approximated as spherical waves (see Fig.~\ref{fig:model} (b)). Otherwise, it is considered as the far-field channel and the wavefronts are approximated as planar waves (see Fig.~\ref{fig:model} (a)). Using the array response vectors in Definition~\ref{def:array_vector}, $\{\Hm_k:k\in[K]\}$ can be described as follows:
\begin{itemize}
    \item Near-field channel ($z_{h_k} \leq Z_{\rm RD}$):
\begin{align*}
    \Hm_k &= \sum_{i=1}^{N_{h_k}}\alpha_{h_k}^{i}\bv_{N}\left(d_{[h_k,{\rm r}]}^{i},\theta_{[h_k,{\rm r}]}^{i},\phi_{[h_k,{\rm r}]}^{i}\right)\nonumber\\
    &\quad\quad\quad\quad\quad\quad\quad\quad\quad\quad     \times \av_{L}^{\herm}\left(\theta_{[h_k,{\rm t}]}^{i},\phi_{[h_k,{\rm t}]}^{i}\right),
\end{align*} 
    \item Far-field channel ($z_{h_k} > Z_{\rm RD}$)
\begin{align*}
    \Hm_k = \sum_{i=1}^{N_{h_k}}\alpha_{h_k}^{i}\av_{N}\left(\theta_{[h_k,{\rm r}]}^{i}\phi_{[h_k,{\rm r}]}^{i}\right)\av_{L}^{\herm}\left(\theta_{[h_k,{\rm t}]}^{i}\phi_{[h_k,{\rm t}]}^{i}\right),
\end{align*} where $N_{h_k}$ is the number of spatial paths, $d_{[h_k,{\rm r}]}^{i}$ is the distance between $i$-th scatter and the RIS, $\theta_{[h_k,{\rm r}]}^{i}$ (resp. $\theta_{[h_k,{\rm t}]}^{i}$) and $\phi_{[h_k,{\rm r}]}^{i}$ (resp. $\phi_{[h_k,{\rm r}]}^{i}$) are the elevation and azimuth angle of arrival (resp. departure) for the RIS (resp. user $k$), respectively.
\end{itemize}

%the MIMO Rayleigh distance (MIMO-ARD), denoted by $Z_{\rm RD}$, and MIMO advanced Rayleigh distance (MIMO-ARD), denoted by  $Z_{\rm ARD}$. The former determines the boundary between the near-field XL-MIMO channel and the far-field MIMO channel (see (d) and (e) in Fig. 2) and the latter determines the boundary between the mixed LoS/NLoS near-field XL-MIMO channel and the conventional near-field XL-MIMO channel in which the LoS component can be modeled by the multiplication of near-field steering vectors at the RIS and the BS (see (c) and (d) in Fig. 2).

%%%%%%%%%%%%%%%%%%%%%%%%%%%
\subsection{BS-RIS Channel}\label{subsubsec:BS-RIS}

We describe the BS-RIS channel (i.e., $\Fm$). To classify this channel, we define the MIMO Rayleigh distance (MIMO-RD), given by \cite{Lu2023}:
\begin{equation}
   Z_{\rm MRD} \eqdef \frac{2(D_{\rm RIS} + D_{\rm BS})^2}{\lambda},
\end{equation} where $D_{\rm RIS}=(\lambda/2)N$ and $D_{\rm BS} = (\lambda/2)M$ denote the antenna apertures of the RIS and the BS, respectively. Using the array response vectors in Definition~\ref{def:array_vector}, $\Fm$ can be described as follows:
\begin{itemize}
    \item Near-field channel ($z_{f} \leq Z_{\rm MRD}$):
\begin{align*}
    \Fm &= \sum_{i=1}^{N_f}\alpha_{f}^{j}\bv_{M}\left(d_{[f,{\rm r}]}^{i},\theta_{[f,{\rm r}]}^{i},\phi_{[f,{\rm r}]}^{i}\right)\nonumber\\
    &\quad\quad\quad\quad\quad\quad\quad\quad\quad\quad 
 \times \bv_{N}^{\herm}\left(d_{[f,{\rm t}]}^{i},\theta_{[f,{\rm t}]}^{i},\phi_{[f,{\rm t}]}^{i}\right),
\end{align*} 
    \item Far-field channel ($z_{f} > Z_{\rm MRD}$):
\begin{align*}
    \Fm = \sum_{i=1}^{N_f}\alpha_f^{i}\av_{M}\left(\theta_{[f,{\rm r}]}^{i},\phi_{[f,{\rm r}]}^{i}\right)\av_{N}^{\herm}\left(\theta_{[f,{\rm t}]}^{i},\phi_{[f,{\rm t}]}^{i}\right),
\end{align*}
where $N_{f}$ is the number of spatial paths, $d_{[f,{\rm r}]}^{i}$ (resp. $d_{[f,{\rm t}]}^{i}$) are distance between $i$-th scatter and the BS (resp. the RIS), $\theta_{[f,{\rm r}]}^{i}$ (resp. $\theta_{[f,{\rm t}]}^{i}$) and $\phi_{[f,{\rm r}]}^{i}$ (resp. $\phi_{[f,{\rm t}]}^{i}$) are elevation and azimuth angle of arrival (resp. departure) for the BS (resp. the RIS), respectively.
\end{itemize}

For the first time, we propose an {\em unified} channel estimation method in Section~\ref{sec:method}, which can perform in the following categories of wireless channels:
\begin{itemize}
    \item Far-far field channels: Both $\Fm$ and $\{\Hm_{k}: k \in [K]\}$ are all far-field channels.
    \item Far-near field channels: $\Fm$ is a far-field channel and $\{\Hm_{k}: k \in [K]\}$ are all near-field channels.
    \item Near-near field channels: Both $\Fm$ and $\{\Hm_{k}: k \in [K]\}$ are all near-field channels.
\end{itemize} On the other hand, in the existing compressed sensing (CS)-based methods \cite{chen2023channel, Yang2023}, a dictionary should be designed by taking into account the characteristics of the near- and far-field channels.

\begin{figure*}[t]
\centering
\includegraphics[width=1.0\linewidth]{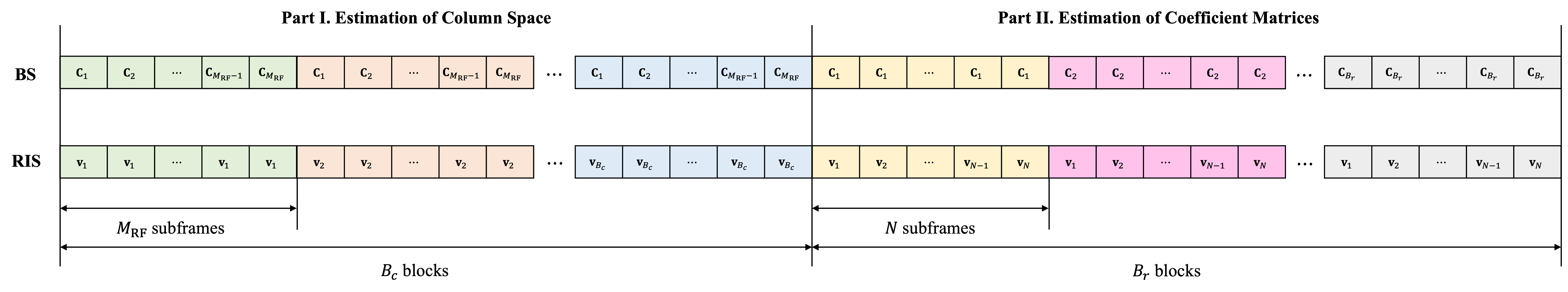}
\caption{The proposed channel estimation protocol and frame structure, where each subframe consists of $T$ (i.e., the length of a pilot sequence) time slots, where $M_{\rm RF}=M/N_{\rm RF}$.}
\label{fig:protocol}
\end{figure*}
%%%%%%%%%%%%%%%%%%%%%%%%%%%%%%%%%%%%%%%%%%

%%%%%%%%%%%%%%%%%%%%%%%%%%%%%%%%%%%%%%%%%%%%%%%%%%%%%%%%%%%%%%%%%%%%%%%%%
%
%%%%%%%%%%%%%%%%%%%%%%%%%%%%%%%%%%%%%%%%%%%%%%%%%%%%%%%%%%%%%%%%%%%%%%%%%
\section{Channel Estimation Protocol}\label{subsec:Channel estimation}

The channel estimation protocol performs with $J$ subframes, each of which consists of $T$ symbols (or time slots) with $T \geq KL$. Accordingly, it requires the total $J T$ symbols, where the hyperparameter $J$ can control the tradeoff between the estimation accuracy and the training overhead. Throughout the paper, $J$ is referred to as the training overhead. In the proposed protocol, the RIS reflection vector $\vv_{j}=[v_{[j,1]},v_{[j,2]},...,v_{[j,N]}]^{\intercal}$ and the $M\times N_{\rm RF}$ RF combining matrix $\Cm_{j}$ are unchanged within each subframe $j \in [J]$, but they can be changed across the subframes. Their specific constructions will be provided in Section~\ref{sec:method}.

During the $J$ subframes, our channel estimation protocol proceeds as follows. At every subframe $j \in [J]$, each user $k \in [K]$ transmits its orthogonal pilot sequence of the length $T$ to the BS, denoted by $\Xm_{[j,k]} \in \CC^{L\times T}$, such that 
\begin{equation}
    \Xm_{[j,k]}\Xm_{[j,k]}^{\herm} =PT\times \Id, \label{eq:power}
\end{equation} where $PT$ is the energy constraint, $P$ is the transmit power of each user and antenna, and $\Xm_{[j,k]}\Xm_{[j,k']}^{\herm} = {\bf 0}$ if $k\neq k'$. This construction is always possible due to the choice of $T\geq KL$. Note that each user $k$ sends the $i$-th row of $\Xm_{[j,k]}$ through the $i$-th transmit antenna during the $T$ time slots. For each subframe $j \in [J]$, the BS then observes the $N_{\rm RF}\times T$ matrix, given by
\begin{align}
\Ym_{j} = \Cm_{j}^{\herm}\left(\sum_{k=1}^{K}\left(\Fm {\rm diag}(\vv_j)\Hm_{k}\right)\Xm_{[j,k]} + \Um_{j}\right),\label{eq:variance}
\end{align}  where $\Um_{j}$ is the noise matrix whose elements follow independently identically circularly symmetric complex Gaussian distribution with mean zero and variance $\sigma^2$. Leveraging the orthogonality of the pilot sequences, we can derive the user-specific observation:
\begin{align}
    \Zm_{[j,k]}&=\frac{1}{PT}\Ym_{j}\Xm_{[j,k]}^{\herm} \in \CC^{N_{\rm RF}\times L} \nonumber\\
&=\Cm_{j}^{\herm}\left(\Fm {\rm diag}(\vv_j)\Hm_{k} + {\Nm}_{[j,k]}\right),\label{eq:signal-model}
\end{align} for $k\in [K]$, where ${\Nm}_{[j,k]}\eqdef \frac{1}{PT}\Um_{j}\Xm_{[j,k]}^{\herm}$.
%\begin{equation}
%    {\Nm}_{[j,k]}\eqdef \frac{1}{P}\Um_{j}\Xm_{[j,k]}^{\herm}.
%\end{equation}
Stacking the $K$ user-specific observations, we define:
\begin{align}
\Zm_{j}&\eqdef \begin{bmatrix}
    \Zm_{[j,1]} & \cdots & \Zm_{[j,K]}
\end{bmatrix}\nonumber\\
&=\Cm_{j}^{\herm}\left(\Fm {\rm diag}(\vv_j)\Hm + {\Nm}_{j}\right),\; j \in [J],\label{eq:obs}
\end{align}  where the cascaded channel and noise matrices are respectively defined as 
\begin{align*}
%\Gm^{\herm} &\eqdef \begin{bmatrix}
%    \Gm_{1}^{\herm} & \cdots & \Gm_{K}^{\herm}
%\end{bmatrix}\in \CC^{M\times KL}\nonumber\\
\Hm &\eqdef \begin{bmatrix}
    \Hm_{1} & \cdots & \Hm_{K}
\end{bmatrix}\in \CC^{N\times KL}\\
{\Nm}_{j} &\eqdef \begin{bmatrix}
    {\Nm}_{[j,1]} & \cdots & {\Nm}_{[j,K]}
\end{bmatrix}\in \CC^{N\times KL}.
\end{align*} 

%%%%%%%%%%%%%%%%%%%%%%%%%%%%%%%%%%%%%%%%%%%%%%
\begin{remark}\label{rem:hybrid} 
Due to the use of hybrid beamforming structures, the BS can observe the $N_{\rm RF} \times L$ matrix during each subframe. Whereas, the $M \times L$ matrix is observed when (conventional) fully-digital beamforming is employed. To attain the same amount of the training observations, the training overhead becomes the $M/N_{\rm RF}$ times bigger. Naturally, in our simulations in Section~\ref{sec:SR}, the scale of the training overhead tends to be larger than those in \cite{tsai2018efficient, chen2023channel,Hu2021} based on fully-digital systems. \hfill$\blacksquare$
\end{remark}

%%%%%%%%%%%%%%%%%%%%%%%%%%%%%%%%%%%%%%%%%%%%%%%%%%%%%%%%%%%%%%%%%%%%
\section{The Proposed Channel Estimation Method}\label{sec:method}
%%%%%%%%%%%%%%%%%%%%%%%%%%%%%%%%%%%%%%%%%%%%%%%%%%%%%%%%%%%%%%%%%%%%%

In this section, we present an efficient method to estimate the effective (or cascaded) channels $\{\Hm_{[{\rm eff},k]}: k \in [K]\}$ in \eqref{eq:channel_model1} from the processed training observations $\{\Zm_{j}: j \in [J]\}$ in \eqref{eq:obs}. As shown in Section~\ref{subsubsec:BS-RIS}, the BS-RIS channel (i.e., $\Fm$) is expressed as the linear combination of the $N_f$ rank-1 matrices in the both {\em near-} and {\em far-field} channels. Since the rank of $\Fm$ (denoted by ${\rm rk}(\Fm)$) is less than equal to $N_f$, $\Fm$ is a low-rank matrix in mmWave channels. This is one of the key ingredients for the proposed channel estimation method. Let $\Sm_{\rm col}$ denote the $M\times {\rm rk}(\Fm)$ matrix whose columns are the basis of the column space of $\Fm$ (a.k.a. common column space). Regardless of the channel characteristics (i.e., the near- and far-field channels) of $\Fm$, each effective channel $\Hm_{[{\rm eff},k,\ell]}$ in \eqref{eq:channel_model2} can be represented as the {\em low-rank} decomposition:
\begin{equation}
    \Hm_{[{\rm eff},k,\ell]} = \Sm_{\rm col}\Tm_{[k,\ell]},\; k \in [K], \ell \in [L], \label{eq:decompS}
\end{equation} where $\Tm_{[k,\ell]}\in\CC^{{\rm rk}(\Fm) \times N}$ denotes a user and antenna specific coefficient matrix. On the basis of the low-rank decomposition, the proposed channel estimation method consists of the two parts (see Fig.~\ref{fig:protocol}). In the first part, using the training observations from the first $M_{\rm RF}B$ subframes, the column space of $\Fm$ (i.e., $\Sm_{\rm col}$) is estimated by means of a collaborative low-rank approximation.  Then, using the training observations from the remaining $N B_r$ subframes, the coefficient matrices $\Tm_{[k,\ell]}$'s are jointly optimized. Thus, the proposed channel estimation method is named {\bf C}ollaborative {\bf L}ow-{\bf R}ank {\bf A}pproximation and {\bf J}oint {\bf O}ptimization (CLRA-JO). Also, the training overhead of the proposed CLRA-JO is computed as
\begin{equation}
    J=M_{\rm RF}B_c + NB_r,
\end{equation} where $B_c$ and $B_r$ denote the hyperparameters which can control the tradeoff between the estimation accuracy and the training overhead.
%B_c, B_r로 해주세요!

In the subsequent subsections, we will design the RF combining matrices $\{\Cm_j: j \in [J]\}$ and the reflection vectors $\{\vv_j: j \in [J]\}$, in order to derive the processed observations suitable for the proposed channel estimation method. Before describing them, we provide the useful definition below:
\vspace{0.1cm}
\begin{definition}\label{def:2} 
Let $\Phim_{[X,Y]}$ be a $X\times Y$ matrix having orthonormal columns. In the following, we will construct the RF combining matrices and the reflection vectors using the matrix $\Phim_{[X,Y]}$. In practice, a DFT matrix can be adopted as the $\Phim_{[X,Y]}$. \hfill$\blacksquare$
%For example, a DFT matrix can be used for both continuous and discrete phase shifting RIS systems. In the following subsections, the RF combining matrices and the reflection vectors will be constructed using the matrix $\Phim_{[X,Y]}$.
%Noticeably, in discrete phase shift RIS systems, it is assumed that $\Phim_{[X,Y]}$ a $X\times Y$ 2D-DFT matrix.
%orthonormal matrix. We construct the RF combining matrix and RIS reflection vector only using this definition. Remarkably, in discrete phase shifting RIS systems, it is assumed that $\Phim_{[X,Y]}$ a $X\times Y$ 2D-DFT matrix.
\end{definition}
%%%%%%%%%%%%%%%%%%%%%%%%%%%%%%%%%%%%%%%%%%%%%%%%

%%%%%%%%%%%%%%%%%%%%%%%%%%
\subsection{Estimation of Column Space}\label{subsec:DC estimation}

In the first part of training, the total $M_{\rm RF}B_c$ subframes are employed to estimate $\Sm_{\rm col}$ (see Fig.~\ref{fig:protocol}). To design the RF combining matrices and the reflection vectors, the subframes are partitioned into $B_c$ blocks each having $M_{\rm RF}$ subframes.

From the training observations $\{\Zm_{j}: j \in [M_{\rm RF}B_c]\}$ in \eqref{eq:obs}, we derive the observations suitable to estimate $\Sm_{\rm col}$ by properly designing $\{\Cm_j, \vv_j: j \in [M_{\rm RF}B_c]\}$. For the compactness of expressions,  we re-index the training observations $\{\Zm_{j}: j \in [M_{\rm RF}B_c]\}$:
\begin{equation}
    \Zm^{\rm 1st}_{[i,b]} \eqdef \Zm_{M_{\rm RF}(b-1)+i}, \label{eq:obs_col}
\end{equation} for $i \in [M_{\rm RF}]$ and $b \in [B_c]$. We design the RF combining matrices as
\begin{align}
    \Cm_{M_{\rm RF}(b-1)+i} &= \Phim_{[M,M]}(:,[N_{\rm RF}(i-1)+1:N_{\rm RF}i]),\label{eq:RF-combining}
\end{align} for each $i \in [M_{\rm RF}]$ and for every $b \in [B_c]$. As shown in Fig. 3, $\Cm_{i}$ is used at the $i$-th subframe of every block, namely, the subframe-dependent RF-combining matrices are employed. For each block $b \in [B_c]$, the $M_{\rm RF}$ reflection vectors are constructed as follows:
\begin{align}
    \vv_{M_{\rm RF}(b-1)+i} &= \Phim_{[N,B_c]}(:,b),\; \forall i \in [M_{\rm RF}].\nonumber
\end{align}
Focusing on the block $b \in [B_c]$, we will explain how to obtain the column-sampled observations to estimate $\Sm_{\rm col}$. The exactly same procedures are applied to every block. Letting $\Qm_b \eqdef \mbox{diag}(\vv_{M_{\rm RF}(b-1)+i})\Hm= \mbox{diag}(\Phim_{[N,B_c]}(:,b))\Hm$ and with the above constructions, we obtain the following observations from \eqref{eq:obs}:
\begin{align}
    \Zm^{\rm 1st}_{[i,b]} &= \Cm_{M_{\rm RF}(b-1)+i}^{\herm}\Fm\Qm_b + \tilde{\Nm}_{[i,b]},
\end{align} where $\tilde{\Nm}_{[i,b]}$ is denoted as $ \Cm_{M_{\rm RF}(b-1)+i}^{\herm}\Nm_{M_{\rm RF}(b-1)+i}$. For each block $b\in[B_c]$, we can get:
\begin{align}
\tilde{\Zm}_{[{\rm col},b]} &= \Phim_{[M,M]}\begin{bmatrix}
    \Zm^{\rm 1st}_{[1,b]}\\
    .\\
    .\\
    .\\
    \Zm^{\rm 1st}_{[M_{\rm RF},b]}
\end{bmatrix}\in\CC^{M\times KL}\nonumber\\
&\stackrel{(a)}{=}\Phim_{[M,M]}\Phim_{[M,M]}^{\herm}\left(\Fm\Qm_b +\tilde{\Nm}_{[{\rm col},b]}\right)\nonumber\\
&=\Fm\Qm_b +\tilde{\Nm}_{[{\rm col},b]},\label{eq:ob2}
\end{align} where (a) is due to the proposed RF combining matrices in \eqref{eq:RF-combining} and
\begin{align*}
    \tilde{\Nm}_{[{\rm col},b]} = \Phim_{[M,M]}\begin{bmatrix}
    \tilde{\Nm}_{[1,b]}\\
    .\\
    .\\
    .\\
     \tilde{\Nm}_{[M_{\rm RF},b]}
    \end{bmatrix}.
\end{align*} From $\{\tilde{\Zm}_{[{\rm col},b]}:b\in[B_c]\}$, we can obtain the processed observations suitable for estimating $\Sm_{\rm col}$:
\begin{align}
\Mm_{\rm col}&\eqdef\begin{bmatrix}
    \tilde{\Zm}_{[{\rm col},1]} & \cdots & \tilde{\Zm}_{[{\rm col},B_c]}
    \end{bmatrix}\in\CC^{M\times B_c KL}\nonumber\\
    &=\Fm\Qm + \tilde{\Nm}_{\rm col},\label{eq:average_noise}
\end{align} where $\tilde{\Nm}_{\rm col} \eqdef \left[\tilde{\Nm}_{[{\rm col},1]} \dots \tilde{\Nm}_{[{\rm col},B_c]}\right]$ and $\Qm \eqdef \left[\Qm_1 \dots \Qm_{B_c}\right]$.
Noticeably, since $\Qm$ is a full-rank matrix with high-probability (i.e., ${\rm rank}(\Qm)={\rm rank}(\Hm)=N$ with high-probability), the column space of $\Mm_{\rm col}$ can be equivalent to that of $\Fm$, provided that $B$ is larger than ${\rm rk}(\Fm)$ and the impact of noise is very small. 
%%%%%%%%%%%%%%%%%%%%%%%%%%%%%%%%%%%%%%%%

We explain how to estimate $\Sm_{\rm col}$ from the processed observations $\Mm_{\rm col}$ in \eqref{eq:average_noise}. Recall that $\Sm_{\rm col}$ is the $M \times {\rm rk}(\Fm)$ matrix whose columns are the basis of the column space of $\Fm$. To estimate it, we find the eigenvalue decomposition of the covariance matrix of $\Mm_{\rm col}$:
\begin{align}
\Mm_{\rm col}\Mm_{\rm col}^{\herm} = \tilde{\Sm}_{\rm col}\Sigmam_{\rm col}\tilde{\Sm}_{\rm col}^{\herm},
\end{align} where $\tilde{\Sm}_{\rm col}$ is the eigenvectors corresponding to the eigenvalue matrix 
\begin{equation*}
    \Sigmam_{\rm col}= {\rm diag}\left([\lambda_{[{\rm col}, 1]}, \lambda_{[{\rm col}, 2]}, ..., \lambda_{[{\rm col}, M]}]\right),
\end{equation*} where the eigenvalues $\lambda_{[{\rm col}, \ell]}$'s are ordered in a descending order in magnitude. As suggested in \cite{chen2023channel}, the ${\rm rk}(\Fm)$ (i.e., the rank of $\Fm$ (or $\Sm_{\rm col}$)) can be efficiently estimated using the minimum description length (MDL) criterion \cite{wax1985detection} as
\begin{align}
    \widehat{{\rm rk}}(\Fm) &= \argmin_{n \in [M]}\left\{-\log\left(\frac{\prod_{i=n+1}^{M}\lambda_{[{\rm col},i]}^{\frac{1}{M-n}} }{\frac{1}{M-n}\sum_{i=n+1}^{M} \lambda_{[{\rm col},i]}}\right)^{(M-n){B_c}T} \right.\nonumber \\
    &\;\;\;\;\;\;\;\;\;\;\;\;\;\;\;\;\;\;\;\;\;\;\;\;\;\;\;\; \left. +\frac{1}{2} n(2M-n)\log(B_c T) \right\}.\label{eq:Nf}
\end{align} As long as independent columns are sufficiently sampled, the optimization below can yield a good $\widehat{{\rm rk}}(\Fm)$-dimensional column space of $\Fm$ \cite{chae2023column}:
\begin{align}
    \hat{\Sm}_{\rm col} = \argmin_{\Sm\in \CC^{M\times \widehat{{\rm rk}}(\Fm)},\; \Sm^{\herm}\Sm=\Id}\left\|\left(\Id-\Sm\Sm^{\herm}\right)\Mm_{\rm col}\Mm_{\rm col}^{\herm}\right\|_2^2.\label{eq:opt}
\end{align} 
The objective function can measure the distance between the column spaces of $\Sm$ and $\Mm_{\rm col}$ since the difference of $\Sm\Sm^{\herm}\Mm_{\rm col}$ (i.e., the projections of the columns of $\Mm_{\rm col}$ onto the column space of $\Sm$) and $\Mm_{\rm col}$ should be zero if $\Sm$ and $\Mm_{\rm col}$ has the same column space. From Eckart–Young–Mirsky Theorem \cite{eckart1936approximation}, the optimal solution to the above problem is derived as
\begin{equation}
    \hat{\Sm}_{\rm col}=\tilde{\Sm}_{\rm col}(:,[\widehat{{\rm rk}}(\Fm)]).\label{eq:hat_S}
\end{equation} As in \cite{xu2015cur}, the accuracy of the estimated column space can be measured by the Euclidean distance between the BS-RIS channel (i.e., $\Fm$ and the BS-RIS channel projected on the estimated column space (i.e., $\hat{\Sm}_{\rm col}\hat{\Sm}_{\rm col}^{\herm}\Fm$). Also, its upper bound was derived as
\begin{equation}
    \left\|\left(\Id-\hat{\Sm}_{\rm col}\hat{\Sm}_{\rm col}^{\herm}\right)\Fm\right\|_F^2\leq\frac{\delta}{B_cKL},\label{eq:upper}
\end{equation} for some $\delta>0$. We can obtain a {\em multi-user gain} since the upper bound can decrease as $K$ grows. Namely, given a target accuracy level, we can reduce the pilot overhead $B_c$ inversely proportional to $K$.

%{\RED In\cite{xu2015cur}, they proved that for the optimization problem in \eqref{eq:opt}, the number of sampled columns, $q\eqdef B_cKL$, determines the upper bound for the Euclidean distance between the BS-RIS channel, $\Fm$, and the BS-RIS channel projected on the estimated common column space, $\hat{\Sm}_{\rm col}\hat{\Sm}_{\rm col}^{\herm}\Fm$, with very high probability. Specifically, this can be mathematically represented by
%\begin{equation}
%    \left\|\left(\Id-\hat{\Sm}_{\rm col}\hat{\Sm}_{\rm col}^{\herm}\right)\Fm\right\|_F^2\leq\frac{\delta}{q},\label{eq:upper}
%\end{equation} where $\delta$ is given as a constant. That is, the more linearly independent columns are sampled, the smaller the upper bound in \eqref{eq:upper} becomes.}

%%%%%%%%%%%%%%%%%%%%%%%%%%%%%%%%%%%%%%%%%%%%%%%%%%%%%%%%%%%%%%%%%%%%
\subsection{Estimation of Coefficient Matrices}\label{subsec:RIS aided estimation}

We propose an iterative algorithm to estimate the coefficient matrices 
$\{\Tm_{[k,\ell]}: k\in [K], \ell \in [L]\}$ in \eqref{eq:decompS}. We first identify the some relationship among the effective channels due to their special structure in \eqref{eq:channel_model2}:
\begin{align*}
    \Hm_{[{\rm eff},k,\ell]}&=\Fm {\rm diag}(\Hm_{k}(:,\ell))\\
    &=\Fm\left[{\rm diag}(\Hm_{1}(:,1)) {\rm diag}(\Hm_{1}(:,1))^{-1}\right]{\rm diag}(\Hm_{k}(:,\ell))\\
    &=\Fm {\rm diag}(\Hm_{1}(:,1))\Dm_{[k,\ell]} = \Hm_{[{\rm eff},1,1]}\Dm_{[k,\ell]},
\end{align*} where $\Dm_{[k,\ell]}\eqdef {\rm diag}(\Hm_{1}(:,1))^{-1} {\rm diag}(\Hm_{k}(:,\ell))$ is the $N \times N$ diagonal matrix. Since $\Hm_{[{\rm eff},1,1]}=\Sm_{\rm col}\Tm_{[1,1]}$, each coefficient matrix $\Tm_{[k,\ell]}$ can be expressed as
\begin{equation}
    \Tm_{[k,\ell]} = \Tm_{[1,1]}\Dm_{[k,\ell]},\; k\in [K], \ell \in [L]. \label{eq:scaling}
\end{equation}
We thus need to estimate the $\Tm_{[1,1]}$ and the diagonal matrices $\{\Dm_{[k,\ell]}: k \in [K], \ell \in [L]\}$. To design RF combining matrices and reflection vectors, the subframes in the second part are partitioned into $B_r$ blocks each having $N$ subframes (see Fig.~\ref{fig:protocol}). For simplicity, we let $n_0 \eqdef M_{\rm RF}B$ denote the index of the last subframe in the first part. From the training observations $\{\Zm_{j}: j \in [n_0+1:J]$\} in \eqref{eq:obs}, we derive the processed observations suitable to estimate the $\Tm_{[1,1]}$ and $\{\Dm_{[k,\ell]}: k \in [K], \ell \in [L]\}$ by properly designing $\{\Cm_j, \vv_j: j \in [n_0+1:J]\}$. As before, we re-index the training observations $\{\Zm_{j}: j \in [n_0+1:J]$\} such as
\begin{align}
    \Zm^{\rm 2nd}_{[i,b]} &\eqdef\Zm_{n_0 + M_{\rm RF}(b-1)+i},\; b \in [B_r],i\in[N]. \label{eq:zz}
\end{align} For each $b\in[B_r]$ and for every $i\in[N]$, we design the RF combining matrices $\Cm_{n_0 + M_{\rm RF}(b-1)+i}$ as
\begin{equation}
    \Phim_{[M,N_{\rm RF}B_r]}(:,[N_{\rm RF}(b-1)+1:N_{\rm RF}b]). \label{eq:Com2nd}
\end{equation} For each $i \in [N]$ and for every $b \in [B_r]$, the reflection vector is constructed as 
\begin{equation}
    \vv_{n_0+M_{\rm RF}(b-1)+i}=\Phim_{[N,N]}(:,i). \label{eq:RIS2nd}
\end{equation} With the above constructions and from \eqref{eq:obs} and \eqref{eq:zz}, we can define an user- and antenna-wise observation:
\begin{align*}
\Zm_{[i,b,k,\ell]}^{\rm 2nd}&\eqdef\Zm^{\rm 2nd}_{[i,b]}(:,L(k-1)+\ell) \nonumber\\
&\stackrel{(a)}{=} \Cm_{n_0 + M_{\rm RF}(b-1) + i}^{\herm}\Hm_{[{\rm eff},k,\ell]}\Phim_{[N,N]}(:,i) + \tilde{\Nm}_{[i,b,k,\ell]},
\end{align*} where (a) follows from \eqref{eq:channel_model2}, \eqref{eq:signal-model}, and \eqref{eq:RIS2nd}, and where
\begin{equation*}
\tilde{\Nm}_{[i,b,k,\ell]} = \Cm_{n_0 + M_{\rm RF}(b-1) + i}^{\herm}\Nm_{n_0 + M_{\rm RF}(b-1) + i}(:,L(k-1)+\ell).
\end{equation*}
For each block $b\in[B_r]$, we can get:
\begin{align*}
    \bar{\Zm}_{[b,k,\ell]}&\eqdef\begin{bmatrix}
    \Zm_{[1,b,k,\ell]}^{\rm 2nd} & \cdots & \Zm_{[N,b,k,\ell]}^{\rm 2nd}
    \end{bmatrix}\Phim_{[N,N]}^{\herm}\nonumber\\
    &=\Cm_{n_0 + M_{\rm RF}(b-1) + i}^{\herm}\Hm_{[{\rm eff},k,\ell]}\Phim_{[N,N]}\Phim_{[N,N]}^{\herm} + \tilde{\Nm}_{[b,k,\ell]}\nonumber\\
    &=\Cm_{n_0 + M_{\rm RF}(b-1) + i}^{\herm}\Hm_{[{\rm eff},k,\ell]}+\tilde{\Nm}_{[b,k,\ell]},
\end{align*} where 
%the last term comes from the proposed RIS reflection vectors in \eqref{eq:RIS2nd} and 
\begin{equation*}
\tilde{\Nm}_{[b,k,\ell]}=\left[\tilde{\Nm}_{[1,b,k,\ell]}\dots\tilde{\Nm}_{[N,b,k,\ell]}\right]\Phim_{[N,N]}^{\herm}.
\end{equation*} Stacking $\{\bar{\Zm}_{[b,k,\ell]}:b\in[B_r]\}$, we define:
\begin{align}
    \Mm_{[{\rm row},k,\ell]}&\eqdef\begin{bmatrix}
    \bar{\Zm}_{[1,k,\ell]}\\
    .\\
    .\\
    .\\
    \bar{\Zm}_{[B_r,k,\ell]}
    \end{bmatrix}\in\CC^{N_{\rm RF}B_r\times N}\nonumber\\
    &\stackrel{(a)}{=}\Phim_{[M,N_{\rm RF}B_r]}^{\herm}\Hm_{[{\rm eff},k,\ell]} + \tilde{\Nm}_{[{\rm row},k,\ell]}\nonumber\\
    &\stackrel{(b)}{=}\Phim_{[M,N_{\rm RF}B_r]}^{\herm}\Sm_{\rm col}\Tm_{[k,\ell]} + \tilde{\Nm}_{[{\rm row},k,\ell]},\label{eq:rowsamp}
\end{align} where (a) is due to the proposed RF combining matrices in \eqref{eq:Com2nd}, (b) is from the decomposition in \eqref{eq:decompS}, and
\begin{equation*}
    \tilde{\Nm}_{[{\rm row},k,\ell]} \eqdef \begin{bmatrix}
    \tilde{\Nm}_{[1,k,\ell]}\\
        .\\
        .\\
        .\\
\tilde{\Nm}_{[B_{r},k,\ell]}
    \end{bmatrix}.
\end{equation*} Letting $\Pm \eqdef \Phim_{[M,N_{\rm RF}B_r]}^{\herm}\hat{\Sm}_{\rm col}\in\CC^{N_{\rm RF}B_r\times\widehat{{\rm rk}}(\Fm)}$, we directly derive the solution of the individual least-square (LS) estimation using the pseudo-inverse $\Pm^{\dag}$:
\begin{align}%\label{eq:TLS}
    \hat{\Tm}_{[k,\ell]}^{\rm LS} &\eqdef \Pm^{\dag}\Mm_{[{\rm row},k,\ell]}\in\CC^{\widehat{{\rm rk}}(\Fm)\times N}\nonumber\\
    &=\Pm^{\dag}\left(\Phim_{[M,N_{\rm RF}B_r]}^{\herm}\Hm_{[{\rm eff},k,\ell]} + \tilde{\Nm}_{[{\rm row},k,\ell]}\right)\nonumber\\
    &=\Pm^{\dag}\left(\Phim_{[M,N_{\rm RF}B_r]}^{\herm}\Sm_{\rm col}\Tm_{[k,\ell]} + \tilde{\Nm}_{[{\rm row},k,\ell]}\right).\label{eq:TLS}
\end{align} Here, the existence of the
$\Pm^{\dag}$ can be guaranteed by choosing the hyperparameter $B_r$ such that
\begin{equation}
    N_{\rm RF}B_r\geq\widehat{{\rm rk}}(\Fm).
\end{equation} Normally, we have $B_r=1$ for very sparse channel, i.e., $N_{\rm RF}\geq\widehat{{\rm rk}}(\Fm)$.

The LS solutions would be good in very high SNRs. Whereas, it is well-known that the multiplication of the pseudo-inverse would increase the power of the noise vector $\tilde{\Nm}_{[{\rm row},k,\ell]}$, which can degrade the performance in practical SNRs. Thus, there is a room to enhance the estimation accuracy harnessing the property in  \eqref{eq:scaling}.
%As shown in \eqref{eq:TLS}, the similarity between $\hat{\Sm}_{\rm col}$ and $\Sm_{\rm col}$, i.e., the estimation accuracy of \eqref{eq:opt}, and the power of $\tilde{\Nm}_{[{\rm row},k,\ell]}$ affect on the accuracy of above LS estimation. These would be robust to these factors, however there is still a room to enhance the accuracy harnessing the property in \eqref{eq:scaling}.} To address this, we jointly optimize the $\Tm_{[1,1]}$ and $\{\Dm_{[k,\ell]}: k \in [K], \ell \in [L]\}$ by taking the solution of 
\begin{align}
\argmin_{\Tm_{[1,1]},\Dm_{[1,1]},...,\Dm_{[K,L]}}\mathcal{L}&\left(\Tm_{[1,1]}, \{\Dm_{[k,\ell]}\} \right)\nonumber\\
{\rm subject\; to}\; \Dm_{[k,\ell]}'s\;& {\rm are\; diagonal\; matrices}, \label{eq:const}
\end{align} where
\begin{equation*}
    \mathcal{L}\left(\Tm_{[1,1]}, \{\Dm_{[k,\ell]}\} \right)\eqdef\sum_{k=1}^{K}\sum_{\ell=1}^{L}\left \|\hat{\Tm}_{[k,\ell]}^{\rm LS} -\Tm_{[1,1]}\Dm_{[k,\ell]}\right\|_F^2.
\end{equation*}
%where $\Dc = \{\Dm_{1},...,\Dm_{K}\}$. 
We efficiently solve this problem via an alternating optimization (see Algorithm 1).

\vspace{0.2cm}
\noindent{\bf Iterations.} At the $t$-th iteration, the proposed alternating optimization is performed as follows:
\begin{itemize}
    \item For the fixed $\hat{\Tm}_{[1,1]}^{(t-1)}$, our optimization can be formulated as
        \begin{align}
        \hat{\Dm}_{[k,\ell]}^{(t)}&=\argmin_{\Dm_{[k,\ell]}} \Lc\left(\hat{\Tm}_{[1,1]}^{(t-1)},\Dm_{[k,\ell]}\Big|\hat{\Tm}_{[1,1]}^{(t-1)}\right)\nonumber\\
        &=\argmin_{\Dm_{[k,\ell]}}\left\|\hat{\Tm}_{[k,\ell]}^{\rm LS}-\hat{\Tm}_{[1,1]}^{(t-1)}\Dm_{[k,\ell]}\right\|_F^2,\label{eq:it2}
    \end{align} for $k \in [K],\;\ell\in[L]$, where the initial value is determined as $\hat{\Tm}_{[1,1]}^{(0)}=\hat{\Tm}_{[1,1]}^{\rm LS}$. This LS problem is easily solved as
    \begin{equation}
        \hat{\Dm}_{[k,\ell]}^{(t)} = {\rm diag}\left(\Rm(:,\Omega_{N})^{\dagger}\hat{\tv}_{[k,\ell]}^{\rm LS}(\Omega_{N})\right), \label{eq:optDk}
    \end{equation} where $\Rm = \Id \otimes \hat{\Tm}_{[1,1]}^{(t-1)} \in \CC^{N_{f}N \times N^2}$ and
    $\Omega_{N}=\{n+(n-1)N: n \in [N]\}$. Also, vectorization of $\hat{\Tm}_{[k,\ell]}^{\rm LS}$ is denoted as $\hat{\tv}_{[k,\ell]}^{\rm LS} = {\rm vec}(\hat{\Tm}_{[k,\ell]}^{\rm LS}) \in \CC^{\widehat{{\rm rk}}(\Fm) N \times 1}.$
     \item For the fixed $\{\hat{\Dm}_{[k,\ell]}^{(t)}: k\in [K],\ell\in[L]\}$, our optimization can be formulated as a standard LS problem:
    \begin{align}
\hat{\Tm}_{[1,1]}^{(t)}&=\argmin_{\Tm}\mathcal{L}\left(\Tm,\{\hat{\Dm}_{[k,\ell]}^{(t)}\}\Big|\{\hat{\Dm}_{[k,\ell]}^{(t)}\}\right)\nonumber\\
        &=\argmin_{\Tm} \sum_{k=1}^{K}\sum_{\ell=1}^{L}\left\|\hat{\Tm}_{[k,\ell]}^{\rm LS}-\Tm\hat{\Dm}_{[k,\ell]}^{(t)}\right\|_F^2,\label{eq:it1}
    \end{align} where $\hat{\Dm}_{[1,1]}^{(t)}=\Id$ during iterations. The optimal solution is derived as
    \begin{align}
        \hat{\Tm}_{[1,1]}^{(t)} &= \left(\sum_{k=1}^{K}\sum_{\ell=1}^{L}\hat{\Tm}_{[k,\ell]}^{\rm LS}\left(\hat{\Dm}_{[k,\ell]}^{(t)}\right)^{\herm}\right)\nonumber\\
        &\;\;\;\;\;\;\;\;\;\;\;\;\times\left(\sum_{k=1}^{K}\sum_{\ell=1}^{L}\hat{\Dm}_{[k,\ell]}^{(t)}\left(\hat{\Dm}_{[k,\ell]}^{(t)}\right)^{\herm}\right)^{-1}.\label{eq:optT1}
    \end{align}
\end{itemize} 
 After the $t_{\rm max}$ iterations, the estimated cascaded effective channels are given by
\begin{equation}
    \hat{\Hm}_{[{\rm eff},k,\ell]} = \hat{\Sm}_{\rm col}\hat{\Tm}_{[1,1]}^{(t_{\rm max})}\hat{\Dm}_{[k,\ell]}^{(t_{\rm max})}.
\end{equation}

%%%%%%%%%%%%%%%%%%%%%%%%%%%%%%%%%%
% ALGORITHM 1
%%%%%%%%%%%%%%%%%%%%%%%%%%%%%%%%%%
\begin{algorithm}[h]
\caption{Proposed CLRA-JO Algorithm}
\begin{algorithmic}[1]

\State {\bf Input:} $\hat{\Sm}_{\rm col}$ (i.e., the estimated column space of $\Fm$) in \eqref{eq:hat_S}, $\{\Mm_{[{\rm row},k,\ell]}: k\in[K],\ell\in[L]\}$ in \eqref{eq:rowsamp}, and the maximum number of iterations $t_{\rm max}$.
\vspace{0.1cm}
\State {\bf Initialization:} Set $\hat{\Tm}_{[1,1]}^{(0)} = \hat{\Tm}^{\rm LS}_{[1,1]}$ from \eqref{eq:TLS}.
%{\RED
%\vspace{0.1cm}
%\State {\bf Initialization:} 
%\begin{itemize}
%    \item Iteration count $t=0$.
%    \item Compute $\{\hat{\Tm}_{k}^{\rm LS}: k\in[K]\}$ via \eqref{eq:LScoeff}.
%    \item Compute $\{\hat{\Dm}_{k}^{(0)}: k\in[K]\}$ via \eqref{eq:initial}.
%\end{itemize}
%}

\vspace{0.1cm}
\State {\bf Repeat until $t=t_{\rm max}$}
\begin{itemize}
    \item Given $\hat{\Tm}_{[1,1]}^{(t-1)}$, update the $\hat{\Dm}_{[k,\ell]}^{(t)}$ via \eqref{eq:optDk}.
    \item Given  $\{\hat{\Dm}_{[k,\ell]}^{(t)}: k\in[K],\ell\in[L]\}$, update $\hat{\Tm}_{[1,1]}^{(t)}$  via \eqref{eq:optT1}.
\end{itemize}
%\begin{itemize}
%    \item Given $\{\hat{\Dm}_{k}^{(t)}: k\in[K]\}$, update $\hat{\Tm}_{1}^{(t+1)}$  via \eqref{eq:optT1}.
%    \item  Given $\hat{\Tm}_{1}^{(t+1)}$, update $\{\hat{\Dm}_{k}^{(t+1)}: k\in[K]\}$ via \eqref{eq:optDk}.
%    \item Set $t\leftarrow t+1$.
%\end{itemize}
\vspace{0.1cm}
\State {\bf Output:} The estimated RIS-aided effective channels:
\begin{equation*}
    \hat{\Hm}_{[{\rm eff},k,\ell]} = \hat{\Sm}_{\rm col}\hat{\Tm}_{[1,1]}^{(t_{\rm max})}\hat{\Dm}_{[k,\ell]}^{(t_{\rm max})}.
\end{equation*}
\end{algorithmic}
\end{algorithm}

%%%%%%%%%%%%%%%%%%%%%%%%%%%%%%%%%%%%%%
%\begin{figure}[t]
%\centering
%\includegraphics[width=0.8\linewidth]{Sampling.png}
%
%\caption{The description of the partially sampled entries of an RIS-aided effective channel $\Hm_{[{\rm eff},k]}$, where the colored entries indicate the observed samples during the beam training.}
%\end{figure}
%%%%%%%%%%%%%%%%%%%%%%%%%%%%%%%%%%%%%%%%%%

%%%%%%%%%%%%%%%%%%%%%%%%%%%%%%%%%%%%%%%%%%%%%%%%%%%%%%%%%%%
%
%%%%%%%%%%%%%%%%%%%%%%%%%%%%%%%%%%%%%%%%%%%%%%%%%%%%%%%%%%%
\subsection{Analysis}\label{sec:CC}

In this section, we analyze the convergence and the computational complexity of the proposed channel estimation method.

%%%%%%%%%%%%%%%%%%%%%%%%%%%%%%%%%%%%%%%%%%%%%%%%
\subsubsection{Convergence Analysis}\label{sec:CA}

We prove the convergence of the proposed iterative algorithm in Algorithm 1. Toward this, we show that
\begin{equation}
    \mathcal{L}\left(\hat{\Tm}_{[1,1]}^{(t-1)},\left\{\hat{\Dm}_{[k,\ell]}^{(t-1)}\right\}\right) \geq \mathcal{L}\left(\hat{\Tm}_{[1,1]}^{(t)},\left\{\hat{\Dm}_{[k,\ell]}^{(t)}\right\}\right).
\end{equation} This implies that as $t$ grows, the $\hat{\Tm}_{[1,1]}^{(t)}$ and $\{\hat{\Dm}_{[k,\ell}^{(t)}\}$ converge to the optimal solution of our optimization in \eqref{eq:const}. Specifically, the proof is provided as follows:
%\begin{align}
%\mathcal{L}\left(\hat{\Tm}_{1}^{(t-1)},\left\{\hat{\Dm}_{k}^{(t-1)}\right\}\right)&=\mathcal{L}\left(\hat{\Tm}_{1}^{(t-1)},\left\{\hat{\Dm}_{k}^{(t-1)}\right\}\Big|\left\{\hat{\Dm}_{k}^{(t-1)}\right\}\right)\nonumber\\
%&\stackrel{(a)}{\geq}\min_{\Tm_{1}}\;\mathcal{L}\left(\Tm_{1},\left\{\hat{\Dm}_{k}^{(t-1)}\right\}\Big|\left\{\hat{\Dm}_{k}^{(t-1)}\right\}\right)\nonumber\\
%&\stackrel{(a)}{=}\mathcal{L}\left(\hat{\Tm}_{1}^{(t)},\left\{\hat{\Dm}_{k}^{(t-1)}\right\}\Big|\hat{\Tm}_{1}^{(t)}\right)\nonumber\\
%&\stackrel{(a)}{\geq}\sum_{k=1}^{K}\min_{\Dm_{k}}\left\|\hat{\Tm}_{k}^{\rm LS}-\Tm_1^{(t)}\Dm_{k}\right\|_F^2\nonumber\\
%&\stackrel{(b)}{=}\mathcal{L}\left(\hat{\Tm}_{1}^{(t)},\left\{\hat{\Dm}_{k}^{(t)}\right\}\right),
%\end{align} where (a) holds with equality when $\hat{\Tm}_{1}^{(t-1)}$ is an optimum, (b) follows from the fact that $\hat{\Tm}_{1}^{(t)}$ is the optimal solution of the minimization in \eqref{eq:it1}, (c) holds with equality when every $\hat{\Dm}_{k}^{(t-1)}$ is an optimum, and  (d) is due to the fact that $\hat{\Dm}_{k}^{(t)}$ is the optimal solution of the minimization in \eqref{eq:it2}.
\begin{align*}
    &\mathcal{L}\left(\hat{\Tm}_{[1,1]}^{(t-1)},\left\{\hat{\Dm}_{[k,\ell]}^{(t-1)}\right\}\right)=\mathcal{L}\left(\hat{\Tm}_{[1,1]}^{(t-1)},\left\{\hat{\Dm}_{[k,\ell]}^{(t-1)}\right\}\Big|\hat{\Tm}_{[1,1]}^{(t-1)}\right)\nonumber\\
    &\stackrel{(a)}{\geq} \min_{\{\Dm_[k,\ell]\}}\mathcal{L}\left(\hat{\Tm}_{[1,1]}^{(t-1)},\left\{{\Dm}_{[k,\ell]}\right\}\Big|\hat{\Tm}_{[1,1]}^{(t-1)}\right)\nonumber\\
    %&\stackrel{(a)}{\geq}\sum_{k=1}^{K} \min_{\Dm_{K}} \Lc\left(\hat{\Tm}_{1}^{(t-1)},\Dm_k\Big|\hat{\Tm}_{1}^{(t-1)}\right)\nonumber\\
    &\stackrel{(b)}{=}\mathcal{L}\left(\hat{\Tm}_{[1,1]}^{(t-1)},\left\{\hat{\Dm}_{[k,\ell]}^{(t)}\right\}\Big|\hat{\Tm}_{[1,1]}^{(t-1)}\right)\nonumber\\
    &=\mathcal{L}\left(\hat{\Tm}_{[1,1]}^{(t-1)},\left\{\hat{\Dm}_{[k,\ell]}^{(t)}\right\}\right)=\mathcal{L}\left(\hat{\Tm}_{[1,1]}^{(t-1)},\left\{\hat{\Dm}_{[k,\ell]}^{(t)}\right\}\Big|\left\{\hat{\Dm}_{[k,\ell]}^{(t)}\right\}\right)\nonumber\\
    &\stackrel{(c)}{\geq} \min_{\Tm_[1,1]} \mathcal{L}\left({\Tm}_{[1,1]},\left\{\hat{\Dm}_{[k,\ell]}^{(t)}\right\}\Big|\left\{\hat{\Dm}_{[k,\ell]}^{(t)}\right\}\right)\nonumber\\
    &\stackrel{(d)}{=}\mathcal{L}\left(\hat{\Tm}_{[1,1]}^{(t)},\left\{\hat{\Dm}_{[k,\ell]}^{(t)}\right\}\Big|\left\{\hat{\Dm}_{[k,\ell]}^{(t)}\right\}\right)=\mathcal{L}\left(\hat{\Tm}_{[1,1]}^{(t)},\left\{\hat{\Dm}_{[k,\ell]}^{(t)}\right\}\right),
\end{align*} where (a) holds with equality when $\{\hat{\Dm}_{[k,\ell]}^{(t-1)}\}$ are optimum, (b) follows from the fact that each of $\{\hat{\Dm}_{[k,\ell]}^{(t)}\}$ is the optimal solution of the minimization in \eqref{eq:it2}, (c) holds with equality when $\hat{\Tm}_{[1,1]}^{(t-1)}$ is optimum, and (d) is due to the fact that $\hat{\Tm}_{[1,1]}^{(t)}$ is the optimal solution of the minimization in \eqref{eq:it1}.

%% Complexity는 한번 더 꼼꼼히 계산해보겠습니다.
%%%%%%%%%%%%%%%%%%%%%%%%%%%%%%%%%%%%%
\subsubsection{Complexity Analysis}

We derive the computational complexity of the proposed channel estimation method. Following the related work in \cite{chen2023channel}, we measure the computational complexity by counting the number of complex multiplication (CM). Regarding the complexity of the estimation of column space in \eqref{eq:opt}, we need to perform the eigenvalue decomposition, which requires $\mathcal{O}(M^{3})$ computational complexity. Also, the dominant complexity of the proposed joint optimization in Algorithm 1 comes from the computations in \eqref{eq:TLS}, \eqref{eq:optDk} and \eqref{eq:optT1}, which respectively require
\begin{align}
    \Delta_{\rm LS}&\eqdef N\left(\widehat{{\rm rk}}(\Fm)\right)^2KL\\
    \Delta_{D}&\eqdef N(3\widehat{{\rm rk}}(\Fm)+1)KL\label{eq:D}\\
    \Delta_{T}&\eqdef N(\widehat{{\rm rk}}(\Fm) + 2)KL.\label{eq:T}
\end{align} Thus, the overall computational complexity of the proposed CLRA-JO is computed as
\begin{equation}
      \Psi_{\mbox{\scriptsize CLRA-JO}} = \mathcal{O}\left(M^3 + \Delta_{\rm LS} + t_{\rm max}(\Delta_{D}+\Delta_{T})\right).\label{eq:CLRA-complexity}
\end{equation}

\begin{figure}[t]
\centering
\includegraphics[width=0.9\linewidth]{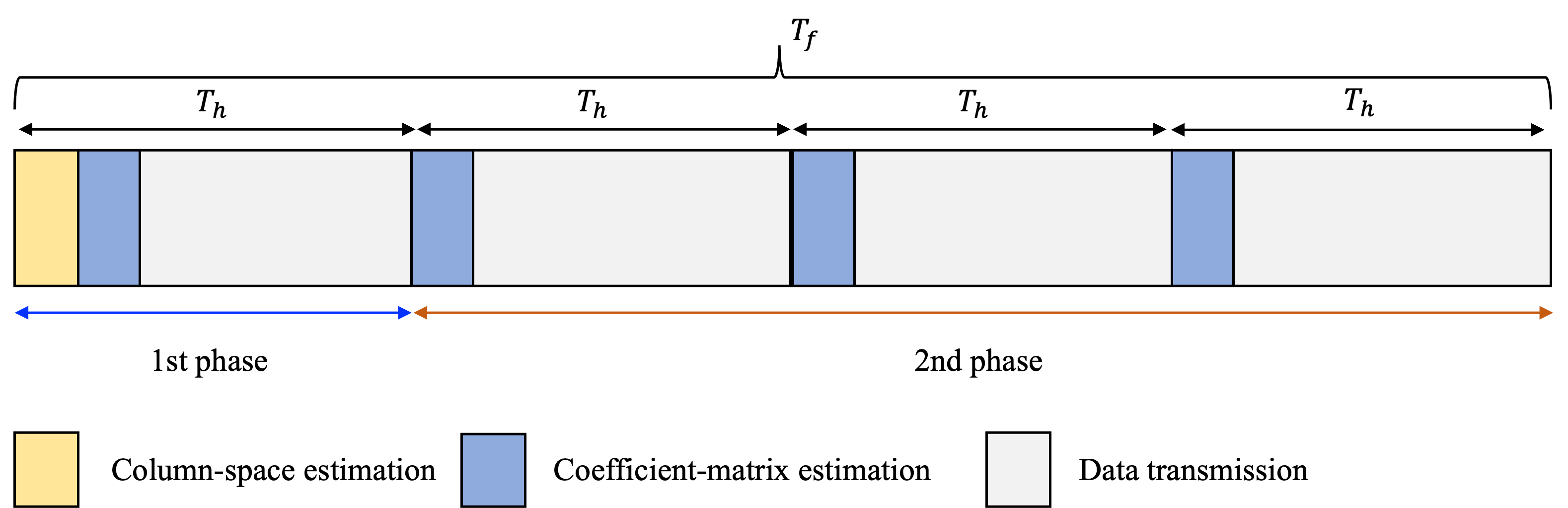}
\caption{The description of two-phase communication protocol.}
\label{fig:two-phase}
\end{figure}

%%%%%%%%%%%%%%%%%%%%%%%%%%
%% 아래 내용 작성 완료되었는지 확인 부탁해요.
%네 작성 완료된 것 같습니다!
\vspace{0.1cm}
\begin{remark}
The proposed channel estimation method can be efficiently operated under the two-phase communication protocol in Fig.~\ref{fig:two-phase}. The key idea is based on the fact that the coherence time of the BS-RIS channel (denoted by $T_f$) is commonly much longer than that of the RIS-User channels (denoted by $T_h$). Accordingly, $\Sm_{\rm col}$ and $\Tm_{[k,\ell]}$'s are changed every $T_f$ and $T_h$ time slots, respectively. Following this protocol, the overall training overhead becomes lower as $T_f$ grows. Namely, in the second phase, we only require the $NB_r$ training overhead, instead of $M_{\rm RF}B_c + N B_r$. In our simulations in Section~\ref{sec:SR}, the effectiveness of the two-phase communication protocol will be demonstrated. Moreover, the two-phase communication protocol can reduce the computational complexity in \eqref{eq:CLRA-complexity} since in this case, the impact of the complexity $\mathcal{O}\left(M^3\right)$ can be negligible. \hfill$\blacksquare$
%In \cite{Hu2021, Yu2023}, the two-phase communication system was proposed to reduce the training overhead. The key idea is to employ the fact that the coherence time of the BS-RIS channel (denoted by $T_f$) is commonly much longer than that of the RIS-User channels (denoted by $T_h$). As depicted in Fig.~\ref{fig:two-phase}, this idea can be applied to the proposed channel estimation method because $\Sm_{\rm col}$ and $\Tm_{[k,\ell]}$'s are changed every $T_f$ and $T_h$ time slots, respectively. Thus, the total training overhead becomes lower as $T_f$ grows, i.e., we need only $NB_r$ training overhead for channel estimation in second phase. In our simulations in Section~\ref{sec:SR}, the effectiveness of the two-phase communication approach will be verified. Definitely, when the two-phase communication approach is employed, the computational complexity of the proposed method in \eqref{eq:CLRA-complexity} can be reduced, i.e., $\mathcal{O}\left(M^3\right)$ is omitted from the complexity in \eqref{eq:CLRA-complexity}. \hfill$\blacksquare$
\end{remark}

%%%%%%%%%%%%%%%%%%%%%%%%%%%%%%%%%%%%%%
%%%%%%%%%%%%%%%%%%%%%%%%%%%%%%%%%%%%%%%%%%

%%%%%%%%%%%%%%%%%%%%%%%%%%%%%%%%%%%%%%%%%%%%
\section{Simulation Results}\label{sec:SR}

In this section, we evaluate the performances of the proposed channel estimation method for RIS-aided mmWave MU-MIMO systems with hybrid beamforming structures. Regarding the wireless channels in our simulations, we consider the XL-RIS assisted XL-MIMO and RIS-aided massive MIMO systems, defined in Section~\ref{subsubsec:BS-RIS}. In Section~\ref{subsec:sim1}, we demonstrate the superiority of the proposed CLRA-JO for XL-RIS assisted XL-MIMO systems by comparing with the state-of-the-art (SOTA) CS-based methods in \cite{chen2023channel}. Remarkably, for the first time, we take into account the near-field BS-RIS and near-field RIS-User channels (in short, near-near field channel). In Section~\ref{subsec:sim2}, we then verify the practicality of the proposed CLRA-JO via experiments on  {\em real} 28GHz UPA channel data in \cite{Akd2014}. Following the performance metric in the related works \cite{tsai2018efficient, chen2023channel, chung2024efficient, JJLee2023}, we employ the normalized mean square error (NMSE) for the evaluation of channel estimation accuracy, given by
\begin{equation}
    \mbox{NMSE} \eqdef \EE\left[\frac{1}{K} \sum_{k=1}^{K} \frac{\|\hat{\Hm}_{[{\rm eff},k]} - \Hm_{[{\rm eff},k]}\|_F^2}{\|\Hm_{[{\rm eff},k]}\|_F^2} \right].
\end{equation} The expectation is evaluated by Monte Carlo simulations with $10^3$ trials.

\subsection{XL-RIS Assisted XL-MIMO Systems}\label{subsec:sim1}

In this section, we verify the superiority of the proposed channel estimation method (named CLRA-JO) for XL-RIS assisted XL-MIMO systems, by evaluating its performances in the following categories of wireless channels:
\begin{itemize}
    \item Far-far field channels: Both $\Fm$ and $\{\Hm_k:k\in[K]\}$ are all far-field channels. Note that this case is identical to the RIS-aided massive MIMO channel.
    \item Far-near field channels: $\Fm$ is a far-field channel and $\{\Hm_k:k\in[K]\}$ are all near-field channels. 
    \item Near-near field channels: Both $\Fm$ and $\{\Hm_k:k\in[K]\}$ are all near-field channels.
\end{itemize} For our simulations, we consider the MU-MIMO system with $M=N=16\times 8=128$, $L=2\times 2=4$, $N_{\rm RF}=8$, the number of spatial paths $N_f=N_{h_k}=3,\;\forall k\in[K]$, the system frequency $f=50 \;{\rm GHz}$, and the wavelength $\lambda=c/f= 0.006{\rm m}$. According to the system wavelength $\lambda$, the RD and MIMO-RD are respectively computed as
\begin{equation*}
    Z_{\rm RD} =49.152{\rm m}\;\mbox{and}\; Z_{\rm MRD} = 196.608{\rm m}.
\end{equation*} The distance between the BS and the RIS are chosen as $z_f=250{\rm m}$ and $z_f=150{\rm m}$ to build the far- and near-field channels, respectively. Also, the distances between the RIS and the $K$ users (denoted by $\{z_{h_k}:k\in[K]\}$) are uniformly distributed over  $[60{\rm m}, 70{\rm m}]$ and $[20{\rm m}, 30{\rm m}]$ to form the far- and near-field channels, respectively. The signal-to-noise ratio ($\SNR$) is defined as
\begin{equation}
    \SNR = 10\log\left(\frac{P}{\sigma^2}\right)[\mbox{dB}],\label{eq:XLpower}
\end{equation} where $P$ and $\sigma^2$ are defined in \eqref{eq:power} and \eqref{eq:variance}, respectively. Specifically, we set that $\sigma^2=1$ and $P$ is described as a relative value of the noise power. Regarding the modeling of wireless channels, the power is equally divided into each signal path, the $i$-th normalized complex channel gains of $\{\Hm_{k}:k\in[K]\}$ and $\Fm$ are respectively determined as
\begin{equation}
    \alpha_{h_k}^{i} = \sqrt{\frac{NL}{N_{h_k}}}\psi_{h_k}^{i}\; \mbox{and}\; \alpha_{f}^{i} = \sqrt{\frac{MN}{N_{f}}}\psi_{f}^{i},
\end{equation} where $\psi_{h_k}^{i}$ and $\psi_{f}^{i}$ are chosen uniformly from $[0,\pi]$. The angle of all AoDs and AoAs are chosen uniformly and randomly from $[0,\pi]$. 

%We note that $0\mbox{dB}$ implies that all elements of a channel matrix and the corresponding noise matrix in \eqref{eq:signal-model} 
%Noticeably, $0\mbox{dB}$ means that all elements of a channel matrix and corresponding noise matrix in \eqref{eq:signal-model} have a same distribution, which is a high noise perturbation environment.

%%%%%%%%%%%%%%%%%%%%%%%%%
\begin{figure}[t]
\centering
\includegraphics[width=0.9\linewidth]{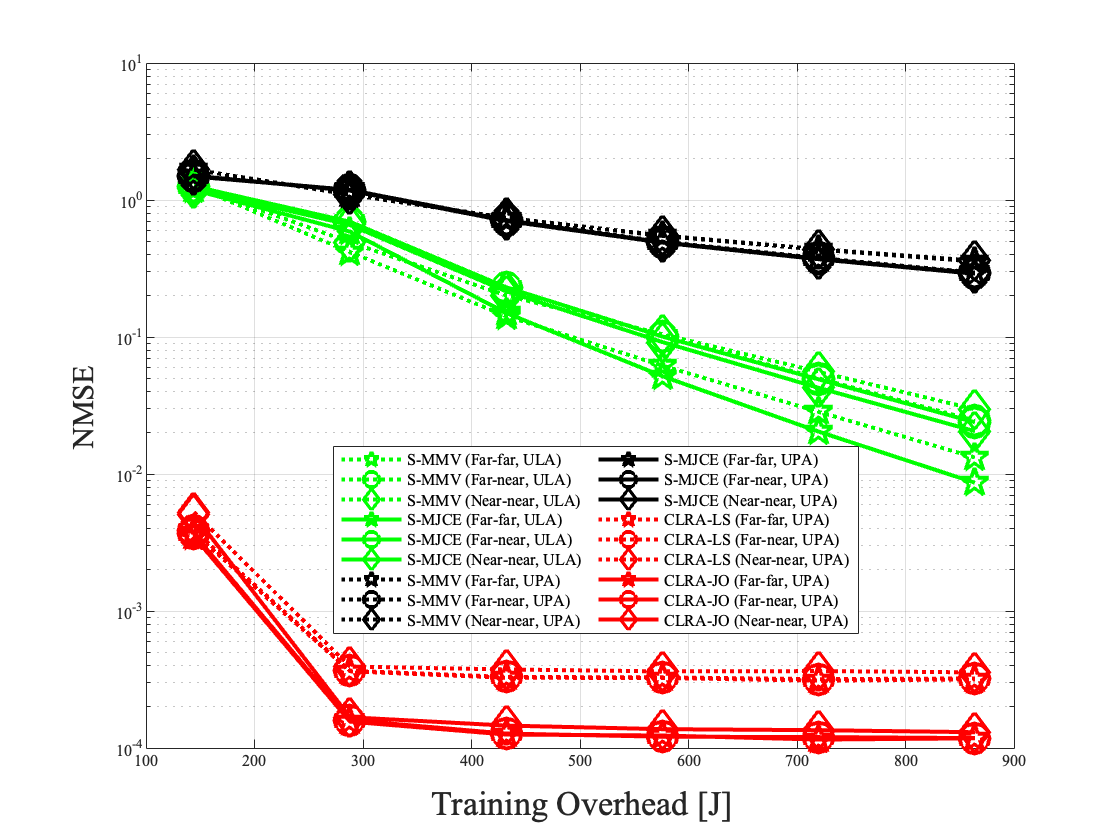}
\caption{The impact of the training overhead on the NMSE.  $\SNR = 0\;\mbox{dB}$ and $K=4$.}
\label{fig:realover}
\end{figure}

\begin{figure}[t]
\centering
\includegraphics[width=0.9\linewidth]{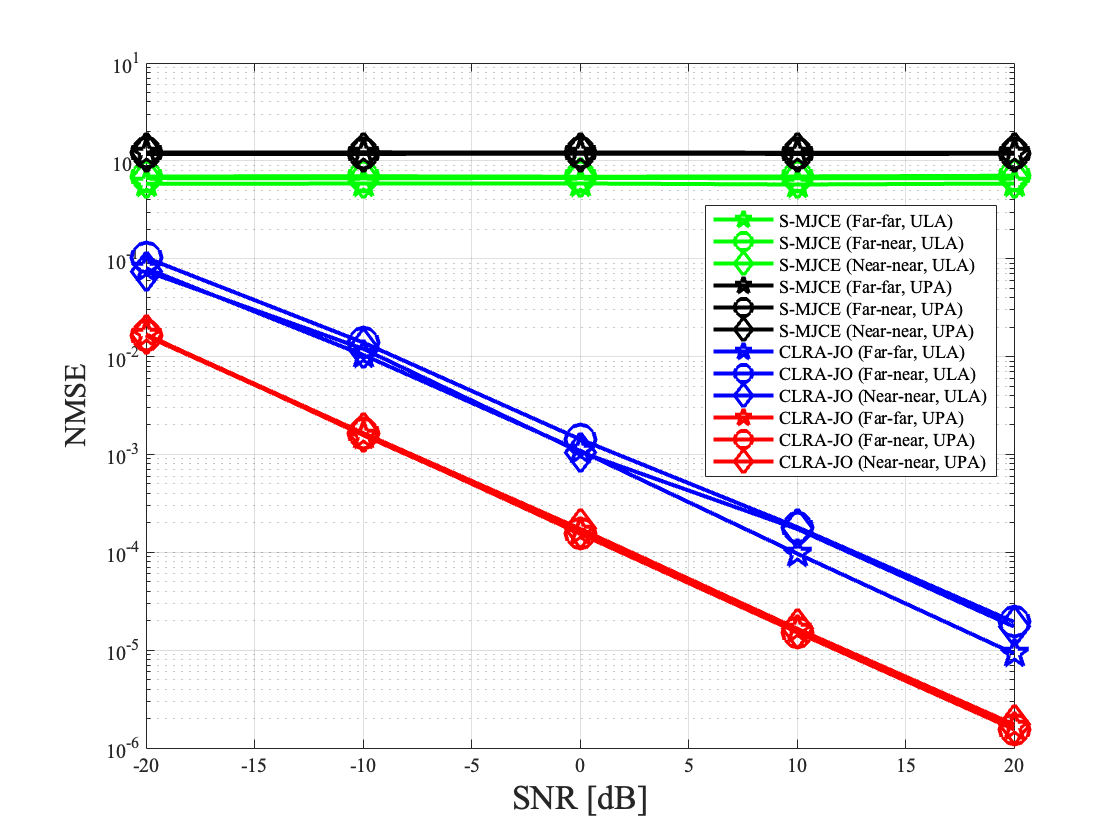}
\caption{The impact of $\SNR$ on the NMSE. $K=4$ and $B_c = 10$.}
\label{fig:realover}
\end{figure}

\begin{figure}[t]
\centering
\includegraphics[width=0.9\linewidth]{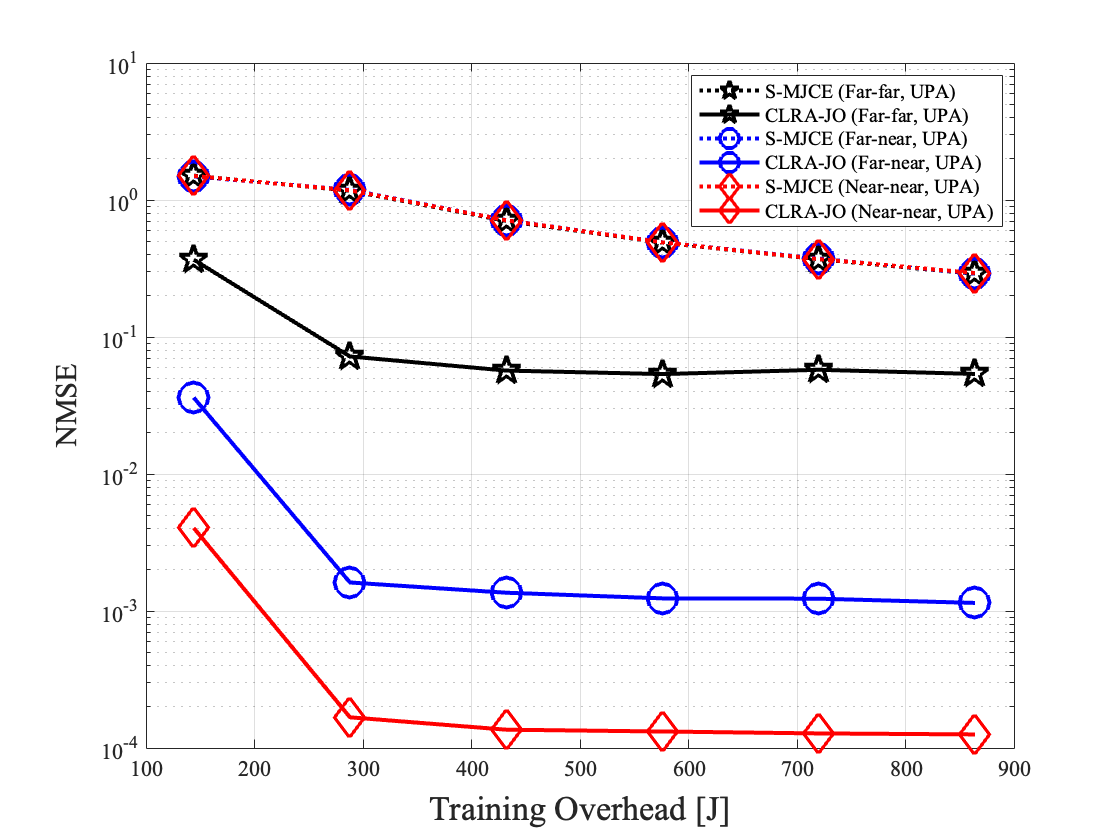}
\caption{The impact of the training overhead and distance on the NMSE. $K=4$.}
\label{fig:realover}
\end{figure}

\begin{figure}[t]
\centering
\includegraphics[width=0.9\linewidth]{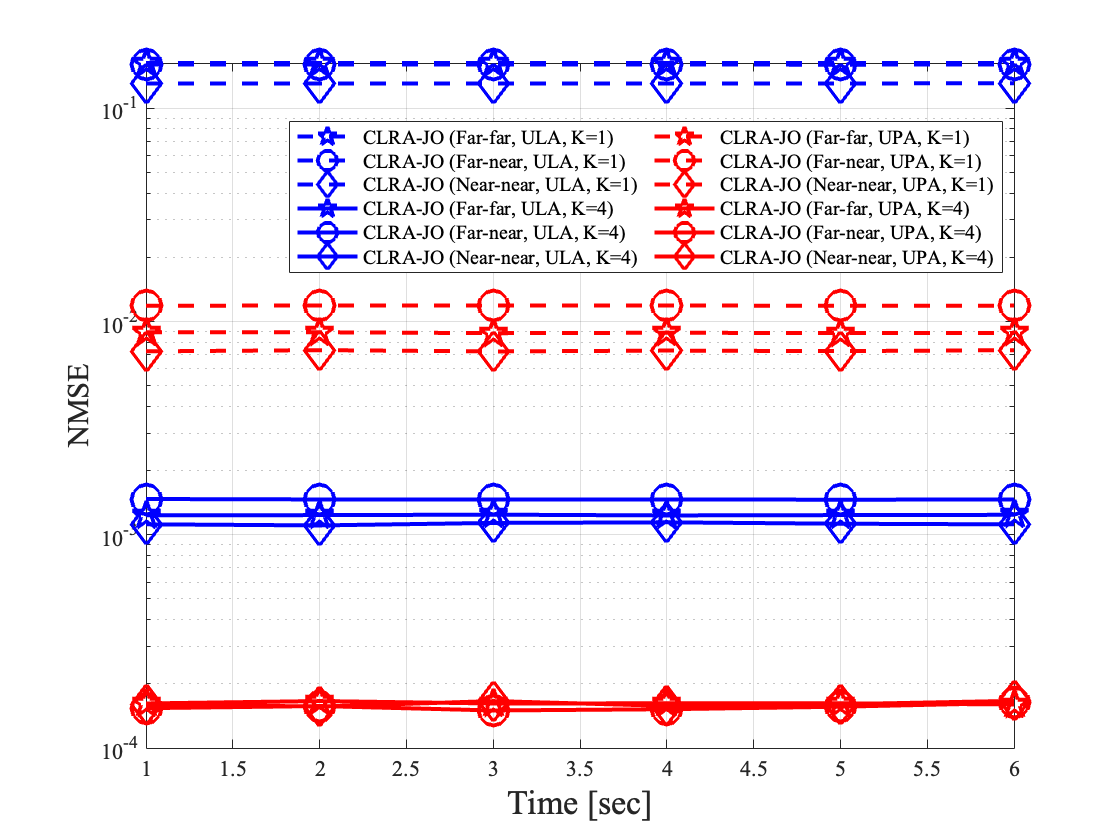}
\caption{The impact of the proposed two-phase communication protocol on the NMSE. $\SNR = 0\;\mbox{dB}$ and $B_c = 10$.}
\label{fig:realover}
\end{figure}

For comparison, the following benchmark methods are considered:
\begin{itemize}
    \item {\bf CLRA-LS}: In the proposed CLRA-JO, the coefficient matrices can be simply obtained from the LS estimation (i.e., $\hat{\Tm}_{k}^{\rm LS}$ in \eqref{eq:TLS}) instead of the joint optimization in Algorithm 1, this simplified method is referred to as CLRA-LS. This method is considered to identify the performance gain of the joint optimization.
    \item {\bf S-MMV}: The channels are estimated by projecting the received signals onto the common column space first, and solved by formulating a multiple measurements vector (MMV) Problem. Here, the MMV problem is addressed via SOMP algorithm. (see Algorithm 1 in \cite{chen2023channel}).
    \item {\bf S-MJCE}: The channels are estimated by the two-step (subspace) multi-user joint channel estimation procedure. (see Algorithm 2 in \cite{chen2023channel}). This method is considered as the best-known CS-based method.
\end{itemize}  
In the benchmark CS methods, the resolution of the quantization for the dictionary are set by $G=512$. Also, in the proposed CLRA-JO, we choose the $B_r = 1$ and $t_{\rm max}=10$ in Algorithm 1.

%수정중!!!!!
\begin{remark} In the existing {\em near-field} channel estimation methods in \cite{Cui2022,Lu2023,Yu2023,Yang2023}, it is assumed that the number of spatial paths for the BS-RIS and RIS-User channels (denoted by $N_f$ and $\{N_{h_k}: k \in [K]\}$, respectively) are exactly known a priori. This makes the existing methods impractical. Also, ULA was only considered since it is quite challenging to design a dictionary representing {\em near-field} UPA array response vectors or solve a 2D-ANM problem in a large system. Thus, they cannot be employed in our system model based on practical UPA and their extension to UPA is non-trivial. Because of this, the CS-based method in \cite{chen2023channel}  was adopted as the benchmark method, wherein UPA dictionaries for both far- and near-field channels are designed by means of the Kronecker product of the horizontal ULA dictionary and the vertical ULA dictionary. This is the common way to design an UPA dictionary \cite{chen2023channel,zheng2023extremely}. \hfill$\blacksquare$
\end{remark}
\vspace{0.1cm}

Fig. 5 shows the impact of training overhead on the NMSE. We observe that the proposed CLRA-based methods can outperform the existing CS-based methods in all the training overhead and all the categories of wireless channels.  Comparing the performances of CLRA-JO and CLRA-LS, we can verify the effectiveness of the proposed joint optimization in Section~\ref{subsec:RIS aided estimation}. Remarkably, the proposed CLRA-JO can attain higher channel estimation accuracy than the SOTA CS-based methods while reducing the pilot overhead about $80\%$ (i.e., $J=144$ and $J=864$ for CLRA-based and CS-based methods, respectively). We also observe that the proposed CLRA-JO shows the almost similar performances regardless of the categories of wireless channels, which verifies that CLRA-JO is indeed the unified method. The CS-based methods perform worse in practical UPA than ULA since they suffer from the severe grid-mismatch problem.

%Also, we observe that the CS-based methods perform worse in practical UPA. This implies that the construction of dictionary representing {\em near-field} array response is demanding.

%Also, the effectiveness of the proposed joint optimization is identified via comparing with CLRA-LS. On the other hand, CS-based methods for ULA channel estimation show better performance than UPAs. This is due to the lack of suitable dictionary representing {\em near-field} array response vectors for RIS/users side.

Fig. 6 shows the impact of $\SNR$ on the NMSE. We first observe that in all the categories of wireless channels, the performances of CLRA-JO improve as $\SNR$ grows. On the other hand, the performances of the CS-based methods are saturated due to the inevitable grid-mismatch problem. Namely, the saturation level is determined according to the quantization resolution of steering vectors. Together with the results in Fig. 5, it is confirmed that the proposed CLRA-JO is more suitable for practical UPA than the CS-based methods.

%내용이 맞는지 검토해주세요.
Fig. 7 shows the impact of training overhead and physical distances on the NMSE. Regarding the BS-RIS and RIS-User channels, the distance-dependent large-scale path loss parameters are given as $PL(z)=10^{-2}z^{-2.2}$ \cite{Rag2019}. While the distribution of the near-near channel and noise remains as in Fig. 5, the channel gains of the far-far and far-near channels are set by the near-near channel and $PL(z)$. We observe that the proposed method can better performance in the near-near channel than the far-near and far-far channels. This is because, in the near-near channel, the $\SNR$ gain of the received signals can be obtained as the physical distance of this channel is shorter than those in the other two channels. Obviously, it is a natural result to be obtained if a channel estimation method is well-developed. In contrast, the CS-based methods cannot attain any estimation gain in terms of distance. 
%Keeping the distribution of the Near-near channel gain and noise the same as Fig. 5, we set the channel gain of the Far-far and the Far-near channel based on the Near-near channel gain and $PL(z)$. We observe that with the same noise power, the estimation performance of the proposed Near-near channel estimation outperforms the performance of both the proposed Far-far channel and the Far-near channel estimation for every training overhead. Whereas, CS-based method doesn't show any estimation gain for distance. From this observation, we can analyze that since the total physical distance of the Near-near field is shorter than others, the power of channel gain of the Near-near channel is larger than others following $PL(z)$, and it leads to the relative growth of SNR under the same power of noise which is determined by the frequency and the bandwidth of the system.  

%As shown in Fig. 5, the proposed CLRA-JO performs well in practical UPA, whereas the CS-based methods do not.

%And, for CLRA-based method, UPA channel estimation shows more accurate estimation than ULAs, whereas CS-based method for UPA channel estimation rather gets worse than UPAs. These imply that our proposed channel estimation is more suitable for practical communication systems than CS-based method. 

%%%%%%%% FIG. 8 %%%%%%%%%
Fig. 8 shows the impact of the proposed two-phase communication protocol in Fig.~\ref{fig:two-phase} on the NMSE. In this simulation, the coherence times are set by  $T_f = 6\;{\rm sec}$ and $T_{h_k} = 1\;{\rm sec},\;\forall k\in[K]$. As expected, the column space estimated only in the first phase (i.e., during the first $1\;{\rm sec}$) can be effectively reused for channel estimations in the remaining second phase. Thus, as aforementioned, this protocol can reduce the computational complexity as well as the training overhead. Under the two-phase communication protocol, the proposed CLRA-JO requires about $2.8\%$ computational complexity of S-MMV (i.e., the CS-based method having the lowest computational complexity). Here, the computational complexity of the S-MMV is provided in \cite{chen2023channel} such as 
\begin{align*}
    \Psi_{\mbox{\scriptsize S-MMV}} &= \mathcal{O}\left(M^3 + N^2(\widehat{{\rm rk}}(\Fm)KL + G) \right).
\end{align*} We also note that the proposed CLRA-JO can indeed yield the multi-user gain since the performance with $K=4$ outperforms that with $K=1$. The performance gain is attained since the accuracy of the common column space estimation and the proposed joint optimization are improved as the $K$ grows.

Based on the above analysis, we can conclude that the proposed CLRA-JO would be a good practical candidate as the channel estimation method for XL-RIS assisted multi-user XL-MIMO.

%Thus, we can conclude that with reference to the channel estimation accuracy about training overhead, SNR and the number of users, we can efficiently design a practical communication system with low training overhead and computational complexity for demanding specification.

%
%%%%%%%%%%%%%%%%%%%%%%%%%%%%%%%%%%%%%%%%%%%%%%%%%%%%%%%%%%%%%%%%%%%%%%%%%%%%%%%%%%
\subsection{RIS-Aided Massive MIMO Systems: 28GHz Channel Data}\label{subsec:sim2}

In this section, we demonstrate the effectiveness of the proposed CLRA-JO for the massive MIMO systems with $M=N=8\times 8 = 64$, $L=2\times 2=4$, and $N_{\rm RF} = 8$. In the proposed method, we choose the $B_r = 1$ and $t_{\rm max}=10$ in Algorithm 1. Also, the {\em real} 28GHz UPA channels are considered, in which  the experimental data in \cite{Akd2014} is used. Regarding the distances among the BS, RIS, and $K$ users, $z_{f}=50{\rm m}$ and $\{z_{h_k}: k\in[K]\}$ are uniformly distributed over $[5{\rm m}, 10{\rm m}]$. Also, the distance-dependent path loss $PL(z)$ ($z=z_{f}\;{\rm or}\; z=z_{h_k}$) is modeled as
\begin{equation}
    PL(z)\;[{\rm dB}] = \alpha + 10\beta\log_{10}(z) + \epsilon,\label{eq:pathloss}
\end{equation} where  $\epsilon \sim \mathcal{N}(0,\nu^{2})$. Here, the channel parameters in \eqref{eq:pathloss} are set by $\alpha = 61.4$, $\beta = 2$, and $\nu=5.8\;{\rm dB}$ for the line-of-sight (LOS) paths, and  $\alpha = 72.0$, $\beta = 2.92$, and $\nu=8.7\;{\rm dB}$ for the non-line-of-sight (NLOS) paths. From these parameters, the complex channel gains of $\{\Hm_k:k\in[K]\}$ and $\Fm$ are independently distributed according to $\mathcal{CN}(0,\gamma_{h_k}^{2}10^{-0.1PL(z_{h_k})})$ and $\mathcal{CN}(0,\gamma_f^{2}10^{-0.1PL(z_f)})$, respectively, where the normalization factor is determined as $\gamma_{h_k} = \sqrt{NL/N_{h_k}}$ and $\gamma_f = \sqrt{MN/N_f}$. In this real experimental data, the $\SNR$ defined in \eqref{eq:XLpower} is defined with 
\begin{equation}
    \sigma^2 = 10^{-0.1(PL(z_{h_k}+PL(z_{f}))}.
\end{equation}

\begin{figure}[t]
\centering
\includegraphics[width=0.9\linewidth]{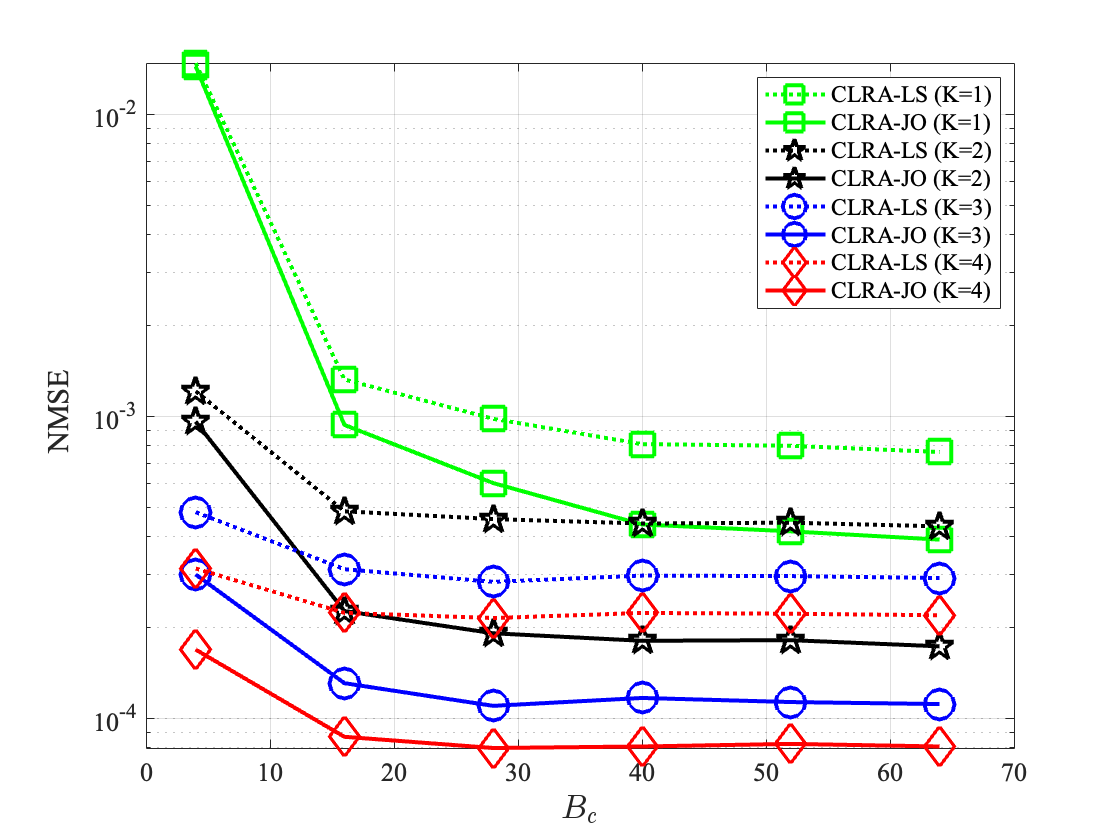}
\caption{The impact of $B_c$ and $K$ on the NMSE. $\SNR = 0\;\mbox{dB}$, $N_f = 3$ and $N_{h_k} = 3,\;\forall k\in[K]$.}
\label{fig:realover}
\end{figure}

\begin{figure}[t]
\centering
\includegraphics[width=0.9\linewidth]{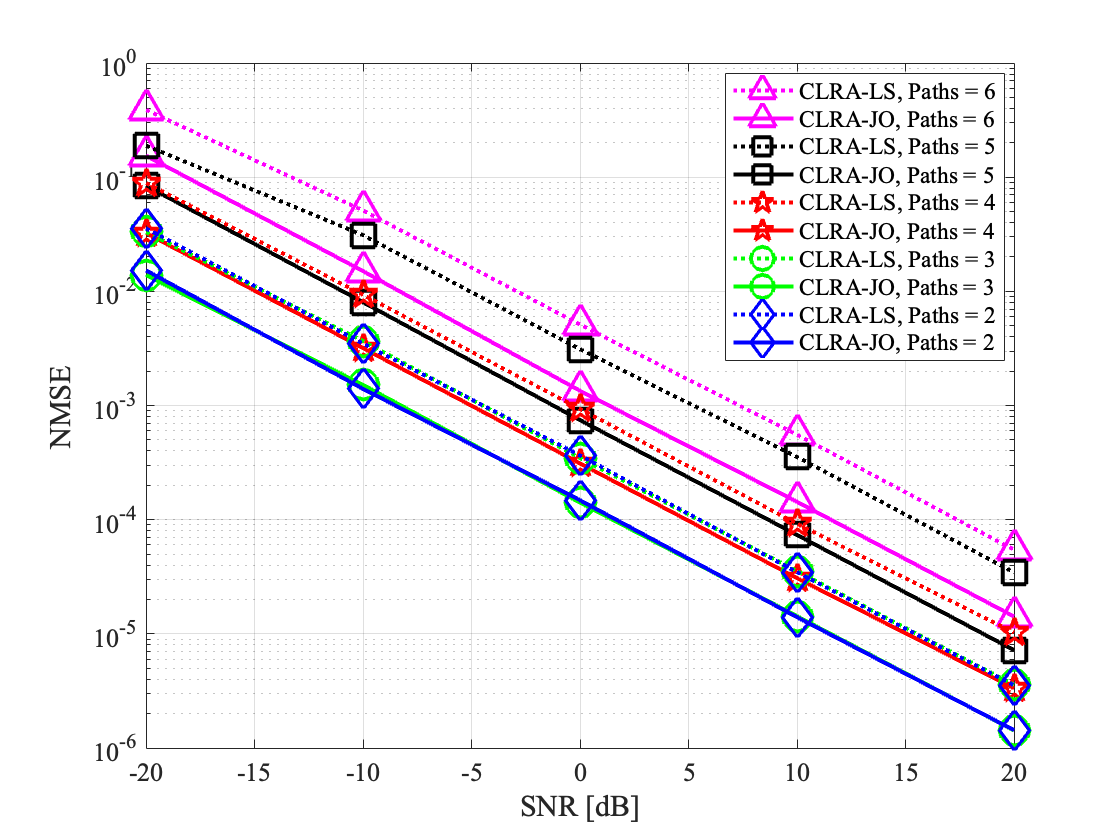}
\caption{The impact of $\SNR$ and $N_{f}$ on the NMSE: $K=4$, $B_c = 10$ and $N_{h_k} = 4\;\forall k\in[K]$.}
\label{fig:real-snr}
\end{figure}

\begin{figure}[t]
\centering
\includegraphics[width=0.9\linewidth]{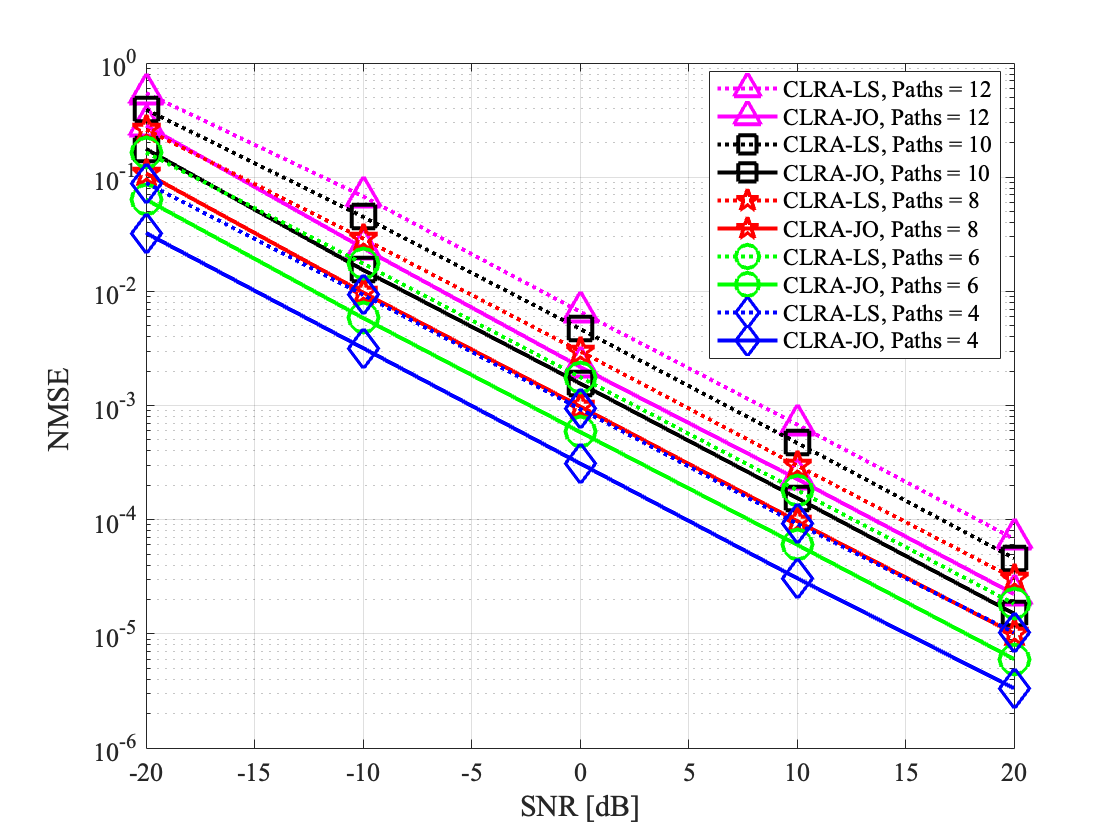}
\caption{The impact of $\SNR$ and $N_{h_k}$, denoted by Paths in label, on the NMSE:  $K=4$, $B_c = 10$ and $N_f = 4$}
\label{fig:real-snrf}
\end{figure}

%%%%%%%%%%%%%%%%%%%%%%%%%%%%%%%%%%%%%%%%%%%%%%%%%%%

Fig. 9 shows the impact of training overhead $B_c$ and the number of users $K$ on the NMSE. From the comparisons with CLRA-LS, we can verify that our joint optimization in \eqref{eq:const}  indeed enhances the estimation accuracy of the user-specific coefficient matrices. We can see that the accuracy of the estimated $\Sm_{\rm col}$ gets better as the number of users $K$ grows. Remarkably, when the number of users is large (i.e., $K\geq 3$), the proposed channel estimation method can achieve an attractive performance with a very small training overhead. Thus, in practical MU-MIMO systems, the proposed method is very promising by considerably reducing the pilot overhead.

%This can be interpreted as the larger the number of columns of $\Mm_{\rm col}$ in \eqref{eq:average_noise} grows, the more accurate the common column subspace $\Sm_{\rm col}$ is estimated. Especially in the large number of users case, i.e., $K=3,4$, the estimation accuracy tends to be constant even with the growth of hyperparemeter $B_c$. This implies that only with very small training overhead, our proposed method can achieve a good estimation accuracy as much as high training overhead.

Figs. 10 and 11 show the impact of SNR and the number of signal paths on the NMSE. As expected, the estimation accuracy is enhanced as $\SNR$ increases (i.e., the noise perturbation in \eqref{eq:TLS} becomes less). From the comparison with CLRA-LS, we can verify the effectiveness of the proposed joint optimization in Section~\ref{subsec:RIS aided estimation}, by alleviating the effect of lower $\SNR$. Specifically, our joint optimization can attain the about $\SNR=5\;\mbox{dB}$ gain compared with CLRA-LS.
%Despite of the $5\;\mbox{dB}$ loss in $\SNR$, specifically, our joint optimization can yield the almost similar estimation accuracy with CLRA-LS.
As expected, the estimation accuracy improves as $N_f$ and $N_{h_k}$ decrease. Thus, the proposed method is well-suited to the next-generation mmWave or THz communication systems, where the corresponding wireless channels consist of lower scattering and a smaller number of signal paths.

%Also, we observe that the estimation accuracy increases as $N_f$ and $N_{h_k}$ decrease. As a expected result, we can conclude that our proposed method is well suited to mmWave systems, where communications are generally performed in environments with low scattering and small number of paths.  

%%%%%%%%%%%%%%%%%%%%%%%%%%%%%%%%%%%%%%%%%%%%%%%%%%%%%%%%%%%%%%%%%%%%%%%%%%%%%%5
%%%%%%%% BENCHMARK METHOD가 없는 이유 설명하고, 아래 CLRA-LS내용 적으면 될 것 같습니다!!!
%%%%%%%%%%%%%%%%%%%%%%%%%%%%%%%%%%%%%%%%%%%%%%%%%%%%%%%%%%%%%%%%%%%%%%%%%%%%%%%%

%%%%%%%%%%%%%%%%%%%%%%%%%%%%%%%%%%%%%%%%%%%%%%%%
\section{Conclusion}\label{sec:C}

We studied the channel estimation problem for XL-RIS assisted multi-user XL-MIMO systems with hybrid beamforming structures. In this system, we proposed the unified channel estimation method (named CLRA-JO) which can perform in the far- and near-field channels without any modification. Whereas, in the existing CS-based methods, dictionary should be designed by taking into account the characteristics of the near- and far-field channels. Via simulations and complexity analysis, it is demonstrated that the proposed CLRA-JO can yield better estimation accuracy than the state-of-the-art CS-based methods while having lower training overhead (e.g., the $80\%$ reduction of the pilot overhead). Our on-going work is to design channel state information (CSI) feedback suitable for the proposed CLRA JO so that it can be applicable in frequency-division-duplexing (FDD)-based XL-RIS assisted multi-user XL-MIMO systems.

\end{document}